# Two base rates, two weights: base-rate neglect has a second axis

A. Y. Shavit · Hunter College and the Graduate Center, CUNY · as1127@hunter.cuny.edu · ORCID: 0009-0008-1235-0995

**Keywords:** base-rate neglect; cue-density effect; contingency learning; signal detection theory; Rescorla–Wagner; Bayesian inference; identifiability

## Abstract

Base-rate neglect is usually treated as one mistake: giving the prior too little weight. Turning co-occurrences into a judgment, though, takes two base-rate corrections: weighting in the prior, how common the outcome is, and dividing out how common the *cue* is. A learner can make either mistake alone. Under-correcting the prior is classical base-rate neglect. Under-correcting the cue is the *cue-density effect* of contingency learning (learning which cues predict outcomes), long studied but not usually framed as the cue-side analogue of base-rate under-correction. We write both corrections as two weights in one Bayesian equation. A leading model of covariation judgment, the dual-factor heuristic, sits where both weights equal one-half. Here the two weights are free.

Response format fixes which weight a study can measure. A graded rating carries the cue-frequency weight directly. A forced choice between two outcomes under one cue cancels it. It measures instead a separate weight in *odds* space, the same correction applied to the posterior odds rather than the absolute judged magnitude; the two weights coincide only at full correction. The cancellation is complete, so the choice proportions of such a task carry no information about the cue-frequency weight, not even its direction. A forced choice between two *cues* keeps the cue-frequency term. In a design meeting a rank condition (cells varied enough to separate parameters), such a choice identifies the weight itself. We show that, given a positive gain and a softmax choice rule, differing cue frequencies across the two alternatives are necessary and sufficient for a choice to depend on the weight.

At their extremes the weights recover base-rate neglect and the prior's share of the signal-detection criterion, the evidence level where responses switch. They also recover the contiguity/sensitivity/validity triple (joint, cue-given-outcome and outcome-given-cue frequency) and lift, the outcome's rate given the cue over its base rate, an association measure that neighbours the causal-strength measures without being one of them. The same cue-frequency weight can be located in, or derived from, six standard learning-and-memory models; in signal detection it appears as its odds-space counterpart. The mappings range from exact identities to limiting or constructed correspondences. In an experiment that keeps only the four cells of a cue × outcome contingency table, each of the six model families, being rich enough to fit them, projects onto one scalar coordinate. That coordinate alone cannot separate accounts free to take any value along it. A rating weight below 1 would count against an account pinned at full correction; signal detection is pinned only in odds space, which ratings do not reach, and whether the exemplar model is pinned is left open. Accounts the coordinate cannot separate may still disagree through their response map, learning trajectory, item memory, or other latent structure.

On a suitable design the two weights are separately identifiable, that is, estimable apart: an experimenter can move the regressor driving either weight without moving the other's, on the outcome side only by holding the cue's frequency fixed. The nested common-weight model, one coefficient for both, excludes that two-coefficient separation by construction. A causal double dissociation would also need interventions that target each normalization's salience alone, not yet run. The separation prediction is the framework's centre, and it has not yet been tested. This paper lays out the framework and the rating experiment that tests that separation. A companion paper fits the odds-space cue weight to a colour–flavour dataset.

---

## 1. Introduction

### 1.1 The standard story: one error, one parameter

For half a century, **base-rate neglect** has been the standard name for a standard failure. If people combine a piece of evidence with the prior probability of a hypothesis — how common that hypothesis is to begin with — they give the evidence too much weight and the prior too little. The prediction studies of Kahneman & Tversky (1973) and the cab problem (Bar-Hillel 1980) are the classic demonstrations. Conservatism (Edwards 1968) is often filed alongside them: revising away from the prior, but not far enough. Behaviourally it is the opposite pattern — insufficient movement *away* from the prior, where neglect is insufficient movement toward it — though that pattern on its own does not settle whether the prior is overweighted or the evidence underweighted (§7 puts both on one axis and returns to exactly this ambiguity). The phenomenon is widely taught as the canonical departure from Bayesian reasoning. Koehler (1996) contests that universality, and Stengård, Juslin, Hahn & van den Berg (2022) test it across a much wider problem space than the low-base-rate corner where demonstrations usually sit. Across measures, it is almost always written down the same way: as a single number that turns the prior down, one underweighting parameter on one term.

The one-parameter picture is tidy, but it assumes there is only one base rate to get wrong. There are two.

### 1.2 The hidden second base rate

Bayes' rule turns the co-occurrences a learner actually experiences into a usable judgment, and it does so by correcting for **two** base rates, not one.

The first is the familiar prior: how common the outcome is. The second is the cue's *own* base rate — how often the cue occurs at all — and correcting for it is a genuinely separate step. A learner sees how often the cue and the outcome go together and, from the outcome's side, how often the outcome comes with the cue. A judgment needs the other direction: given the cue, how often does the outcome follow? Getting from the first to the second means dividing by how often the cue occurs. Skip that division and a cue that merely happens a lot looks more predictive than it is.

A learner can make either correction, both, or neither. Classical base-rate neglect is a failure of the first. This paper is about a failure of the *second*. It is not a new phenomenon: a literature already studies it under another name, without ever calling it base-rate neglect.

### 1.3 The cue-density effect: the same error, measured under another name

In studies of how people judge whether a cue predicts an outcome, two findings recur. A cue that occurs often is judged more strongly associated with the outcome than a rare cue is, even when both predict the outcome equally well (the same validity, $P(O \mid C)$, in Figure 1; in most density studies the same contingency $\Delta P = P(O \mid C) - P(O \mid \neg C)$, often zero) — the **cue-density effect**. And an outcome that occurs often inflates judged strength in the same way — the **outcome-density effect**. Both have been documented since around 1980 (cue-density: Allan & Jenkins 1983; outcome-density: Alloy & Abramson 1979; Shanks & Dickinson 1987) and revisited since (Matute, Yarritu & Vadillo 2011; Matute, Blanco & Díaz-Lago 2019).

In the two-weight parameterization, the cue-density effect *is* the second under-correction: a learner who judges a common cue as stronger than an equally-predictive rare cue, and whose judgment is read through this equation, is under-weighting the cue-frequency correction. The mapping is model-conditional, and §1.5 keeps it so — cell weighting, noise, sampling, pseudocontingency (reading an association off how well the two variables' overall frequencies line up, rather than off the joint cells) and associative accounts can each generate a density effect, so the behavioural effect is not identical to one latent coefficient by definition. But the contingency-learning tradition that measures this effect does not connect it to base-rate neglect in this sense. The nearest exception makes the distinction concrete: Kutzner, Freytag, Vogel & Fiedler (2008) do put base-rate neglect and contingency learning in one title, and find predictions biased toward the more (or less) frequent outcome after the more (or less) frequent cue. The symmetry is the tell. Base-rate *matching* aligns two marginals and so runs in both directions, where a weight on the cue's own frequency pushes a frequent cue's judgments one way whichever outcome is frequent. Theirs is the matching mechanism of §1.5 — the similarity of two base rates taken for a contingency — not a weight on the cue's own frequency in an inversion. And — because these designs vary how often the cue occurs while holding the outcome's base rate fixed — a single such study cannot separate this second neglect from the classical one.

### 1.4 A third reading: the statistical-learning triple

The same structure appears a third time in the statistical-learning treatment of cue–outcome contingency. There the question is whether behaviour tracks the joint co-occurrence, the cue given the outcome, or the outcome given the cue. This paper calls these three readings **contiguity** (here the joint frequency rather than temporal pairing), **sensitivity** and **validity**; the labels are this paper's own, not terms taken from a source. These read the same co-occurrence three ways. Sensitivity and validity share the *same* numerator — the co-occurrence — and differ only in which base rate divides it: sensitivity, $P(C \mid O)$, divides by the outcome's frequency; validity, $P(O \mid C)$, divides by the cue's.

Validity is the one that answers the question the learner faces: given the cue, how likely is the outcome? If *that* is what the task asks for, a learner who tracks sensitivity instead of validity is, again, neglecting the cue's base rate — the qualification matters, because §2 makes which setting counts as correct depend on the target, and sensitivity is not an error where sensitivity is the question. (This is the same mix-up physicians make when they confuse "how often the test fires given the disease" with "how likely the disease is given a positive test" — the inverse fallacy (Eddy 1982); for a demonstration of the error, see Casscells, Schoenberger & Graboys 1978.)

Three literatures, then, one underlying quantity: divide the overlap by the cue circle, or not (Figure 1).

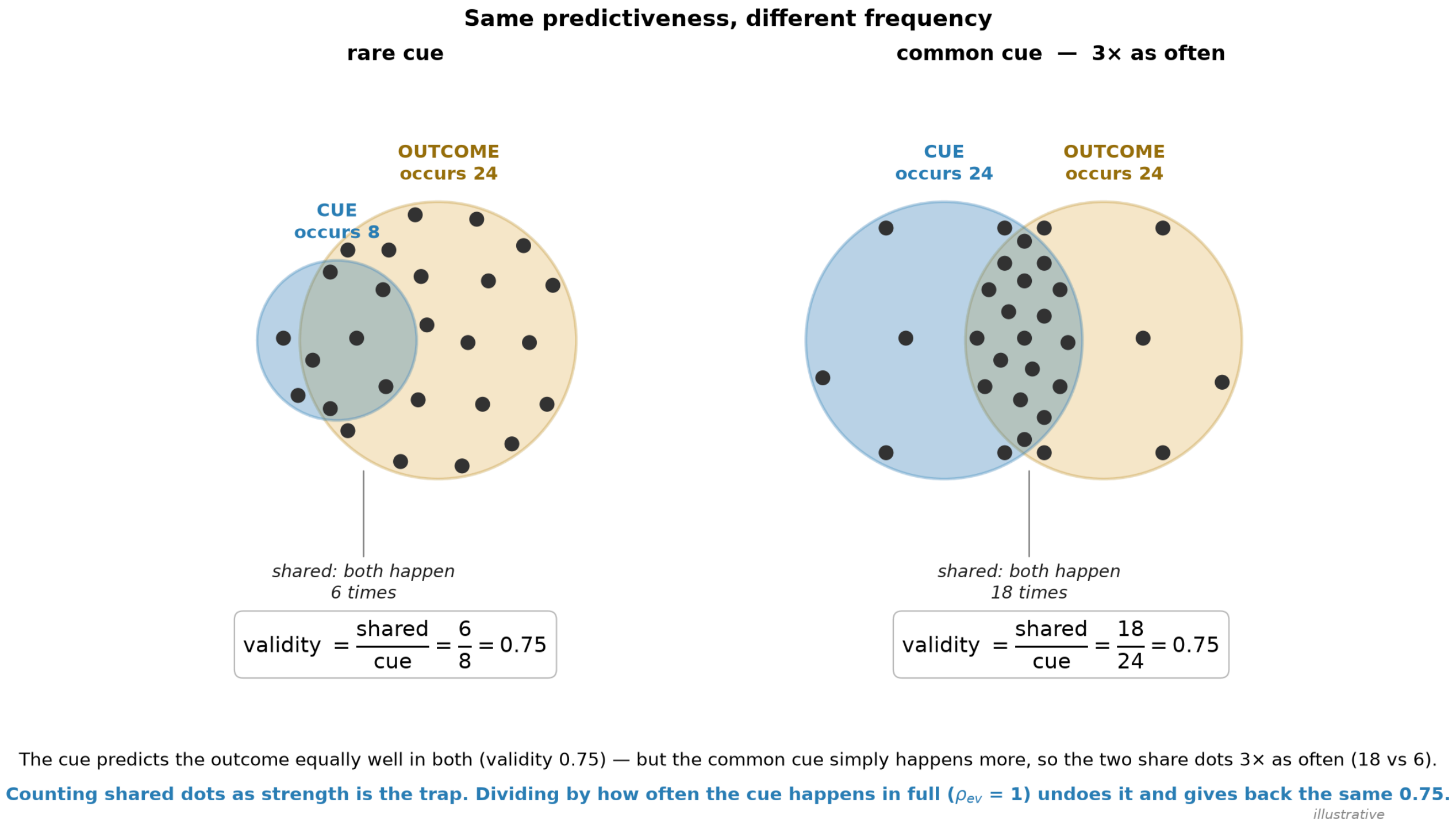

*Figure 1. Same predictiveness, different cue frequency: the confound, concretely.*

Figure 2 sets the three vocabularies side by side: what each field calls the failure (for statistical learning, this paper's labels), how it writes it, and which phenomenon it built its case on. They differ in every surface respect and meet in one equation.

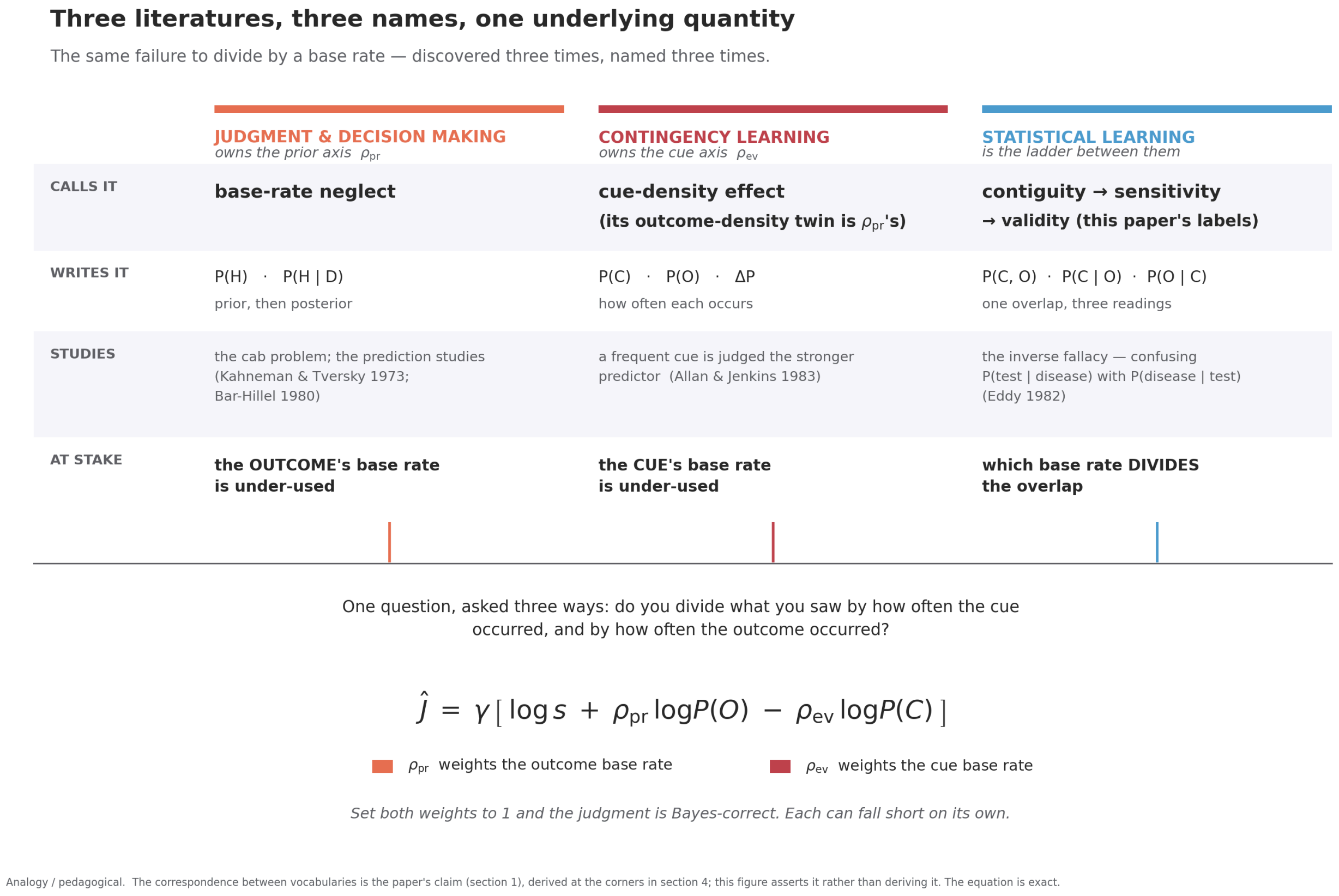

*Figure 2. Three literatures, three vocabularies, one shared quantity.*

## 1.5 The limits of existing accounts

The density tradition has already manipulated both marginals, and even crossed them: Blanco, Matute & Vadillo (2013) varied cue frequency and outcome frequency in one design and asked whether the two biases add up or interact. That tradition does *not* (i) write the two effects as weights in one Bayesian inversion, (ii) connect the cue-density effect to classical base-rate neglect, or (iii) treat the cue-frequency correction as a single coordinate shared across model families. Matute, Blanco & Díaz-Lago (2019) argue that a successful theory of contingency learning should explain accurate *and* biased judgment, and ask whether one common mechanism can do both. A single equation with two weights does exactly that: its calibrated setting is accurate judgment, its under-correcting settings are the biases.

That tradition's own model of the density effects is **cell weighting**: a judgment is fit as a weighted sum of the four cells of the 2×2 table, with the cue-present/outcome-present cell counted most heavily and the doubly-absent cell least (Wasserman et al. 1996). Unequal cell weights reproduce both density effects. The obvious objection is that four cell weights already do the work of two. They do not, but not because cell weights lack a normative setting. They have two. The chi-square logic of testing for any association "would seem to imply that subjects should be equally sensitive to all cells", while a Bayesian comparison of a causal hypothesis against a no-cause hypothesis weights them unequally, so that "every joint occurrence of cause and effect will provide more evidence for the cause than every joint absence" (Anderson & Sheu 1995, pp. 511–512). Joint presence is normatively the more informative cell whenever presence is the rarer state (McKenzie & Mikkelsen 2007), and a cross-validated meta-analytic comparison of covariation models found that a rule weighing confirming against disconfirming evidence fits as well as any rival and best matches the qualitative trends, and proposed a rational justification for it (Perales & Shanks 2007). The difference lies in what each setting is normative *for*. A cell-weight setting is correct relative to a model of the world, an independence test or a causal hypothesis, and it is read against the whole table. The two weights here are anchored to named target quantities, the four corners of §4: 1 applies a correction in full and 0 skips it, so a fitted value says how much of *which* base-rate correction the learner applied, for the question the task asked. The two forms also differ in what they predict. Cell weighting is a statement about a table and is silent on which task can measure what. It does not predict that a two-choice test loses the cue term while a rating keeps it (§2), and it does not predict that cue-side and outcome-side manipulations move two separately identifiable parameters (§3). Those are the two predictions §6 tests.

The rational-heuristic tradition has already placed one model inside the space this paper describes. Hattori & Oaksford (2007) derive the **dual-factor heuristic**, $H = a/\sqrt{(a + b)(a + c)} = \sqrt{P(O \mid C) \cdot P(C \mid O)}$, as the limit of the φ coefficient when the doubly-absent cell grows without bound. Against past data and new experiments it was the best of 41 models of covariation detection. In the notation of §2, H is the geometric mean of validity and sensitivity: $\log H = \log s + \frac{1}{2} \log P(O) - \frac{1}{2} \log P(C)$. That is the two-weight score with $\gamma = 1$ and both weights at one-half (an identity; checked to $9 \times 10^{-16}$ over 100,000 random tables, with controls at other weight settings missing by 0.4 to 1.8). Hattori and Oaksford observe that "H ignores the base rates of the cause and the effect" (p. 770); in this paper's terms, it applies half of each correction. So the leading rational heuristic of covariation detection already sits at the centre of §4's square, on the common-weight line that §6 tests against, midway between the validity and sensitivity corners. This paper does not add that point. It adds the two weights left free and separate, the anchoring of their ends to named targets, and the task dependence of §2, which says which response format can see which weight. A fit of the dual-factor heuristic tests one point on the common-weight line; the design of §6 asks whether learners sit on that line at all.

Two-parameter accounts of misjudgment do already exist. Griffin & Tversky (1992) separate the "strength" of evidence from its "weight"; the generalized-Bayes form (Grether 1980) frees an exponent on the likelihood alongside one on the prior (with subjects classified between prior neglect and prior over-use; El-Gamal & Grether 1995). In both accounts the free parameters attach to quantities our equation already carries, but not to the same ones. Grether's pair sits on the evidence term (which we hold fixed) and on the *prior over the world state*, P(O). Griffin and Tversky's pair sits wholly on the evidence side — how extreme the sample is, their "strength", and how large it is, their "weight" — so neither of their two parameters is a weight on the prior. (Their Study 2 does report the prior underweighted, but as a by-product of strength dominating weight, not as a parameter of its own.) None of the four parameters attaches to the *cue's own base rate*, P(C), which is the coordinate this paper is about. That marginal is exactly the coordinate the generalized-Bayes scaling leaves untouched: it re-weights how much the likelihood and the prior each count, but not how the frequency of the cue itself discounts the read-out. And the single-parameter prior-underweighting model (Benjamin, Bodoh-Creed & Rabin 2019) is exactly the classical one-knob picture.

The closest prior account is **pseudocontingency** (Fiedler, Freytag & Meiser 2009; Fiedler, Kutzner & Vogel 2013), which already gives the cue's base rate first-class explanatory status. If joint observations are unavailable or unusable, a judge reads a contingency off the alignment of the two variables' univariate base rates: the process "relies on the marginal frequencies (base rates) rather than the joint frequencies in the cells of the table" (Fiedler et al. 2009, p. 187). That is the same quantity this paper weights, so the relation needs stating precisely — and it is not rivalry. Pseudocontingency grants the cue marginal its role as a base-rate-*matching heuristic*; it does not write a free weight on log P(C) beside a prior weight in one equation, and it does not recover the lift and contiguity/sensitivity/validity readings as special cases. The factorization instead *derives* the pseudocontingency pattern in the limit where it is measured. At null contingency — the canonical pseudocontingency stimulus — the hit rate s = P(C|O) reduces to P(C), and the judgment collapses to a pure function of the two base rates, with $\rho_{ev} < 1$ as its signature (the companion paper gives the reduction and fits it to published null-contingency data). Pseudocontingency describes *when* base rates drive judgment; the two weights say *which* base-rate term is weighted, and by how much.

Fiedler & Kutzner (2024) press the parsimony case from the other side, arguing that pseudocontingency alone suffices and that two mechanisms are unlikely to run at once. Two weights are not two mechanisms asserted in advance. They are two separately identifiable parameters, and §6's design decides between one and two by fitting both and testing whether they come apart. Parsimony favours the smaller model when the larger fits no better, so it cannot favour either before the comparison exists. The purpose of this paper is to make that comparison possible.

The gap is specific. Having surveyed the base-rate-neglect, contingency-learning and signal-detection literatures cited here, we are unaware of an account that places a free weight on the *cue's base rate* alongside the prior weight (the dual-factor heuristic fixes both at one-half), in one equation that also recovers the signal-detection, lift, and contiguity/sensitivity/validity readings as special cases.

## 1.6 The need, and what this paper offers

Each of the two axes of base-rate neglect is established on its own. Their **unification** is not. Writing both corrections as two weights in one inversion does four things at once that no single existing account does. It names the cue-density effect as a second axis of base-rate neglect. It recovers the classical special cases as corners of one picture: base-rate neglect, the signal-detection criterion, contiguity/sensitivity/validity, and lift, with the dual-factor heuristic of Hattori & Oaksford (2007) at its centre. It shows the two weights are separately identifiable on a suitable design (a proved property of the design, not yet a finding about learners), each estimated from its own regressor while the design holds the other's regressor fixed — a two-coefficient separation the nested common-weight model excludes by construction, and a first step toward the stronger, causal double dissociation that a targeted-salience design would give. And it says which task can measure which weight: the cue-frequency weight cancels exactly from a same-cue two-choice test, a rating carries it directly, and a two-choice test between two *cues* carries it too.

That task dependence is what the framework adds beyond relabelling known effects: it specifies the experiment that would separate them. This paper lays out the framework and that experiment. A companion paper (Shavit, in preparation) fits the cue's odds-space weight to an existing colour–flavour dataset that holds the outcome base rate flat. There the cue's odds-space weight is below its normative value of 1, and identification leans toward the odds-space analogue of sensitivity (the hit rate s, uncorrected) rather than of validity. **That result belongs to the odds-space weight and stops there.** Because a same-cue two-choice task cancels the cue term outright (§2), what such a study measures is a companion weight in odds space, and the cancellation is exact — so the study carries no information about the weight in the equation above, not its size and not its direction. The odds-space finding is worth having on its own terms, as evidence that under-correction of the cue's own likelihood is real in that dataset. It is not a partial indication about the probability-space weight, and we do not offer it as one.

---

## 2. The two-weight model

A learner observes how often a cue and an outcome happen together. A useful judgment needs the reverse of what is easy to count: not "how often is the cue present when the outcome happens," but "given the cue, how often does the outcome follow." Bayes' rule is the step that turns the first into the second, and it does so with two corrections: one for how common the outcome is, one for how common the cue is.

Write the cue's hit rate as $s$ = P(cue | outcome) — how often the cue is present when the outcome occurs — and its false-alarm rate as $s'$ = P(cue | no outcome). Write the outcome's base rate as P(O) and the cue's base rate as P(C). A calibrated learner's judged predictiveness is then the log of the validity P(O | C), times an overall scale factor, the gain $\gamma > 0$ — positive throughout this paper, not merely nonzero, because §3 and §6 read the *sign* of a fitted slope as the direction of a weight, and a negative gain would flip every one of those readings while leaving the equation's fit untouched:

$$\hat{J} = \gamma\,[\,\log s + \log P(O) - \log P(C)\,]$$

Throughout, every marginal lies strictly inside the unit interval. The logarithms are undefined at 0, and the identities below need each of s, s′ and P(O) to be neither impossible nor certain.

The first term is the raw evidence; the second adds in how common the outcome is; the third divides out how common the cue is. Let the learner apply either correction partially, with a weight on each:

$$\hat{J} = \gamma\,[\,\log s + \rho_{pr}\log P(O) - \rho_{ev}\log P(C)\,]$$

$\rho_{pr}$ is how much of the prior correction the learner applies; $\rho_{ev}$ is how much of the cue correction. A weight of 1 applies that correction in full (calibrated); a weight of 0 skips it. **$\rho_{ev}$ < 1 is the cue-density effect**, read through this equation (§1.3), under-dividing by how common the cue is. **$\rho_{pr}$ < 1 is classical base-rate neglect**, under-using how common the outcome is. The two weights sit on two different terms, and (§3) each is estimated from a regressor the design varies on its own. Both weights act on a latent magnitude, not on a rating directly. A rating reads that magnitude through a 0–100 scale, which flattens near each end: §6 turns that curvature into a test, and Appendix A.7 gives its exact form for a logistic link, while Appendix A.4 sets out what the design must satisfy for the three parameters to be identified.

Three weights appear in what follows, and only two of them are free. The third, the evidence weight, first matters in §7. We name it here so that it does not arrive looking like a fourth name for one of the other two:

| weight | corrects for | in this paper |
|---|---|---|
| $\rho_{pr}$, the **prior weight** | how common the outcome is, log P(O) | free; $\rho_{pr}$ < 1 is classical base-rate neglect |
| $\rho_{ev}$, the **cue weight** (in full, the cue-frequency or cue-marginal weight; the subscript does not refer to the evidence term log s, whose weight is fixed at 1) | how common the cue is, log P(C) | free; $\rho_{ev}$ < 1 is the cue-density effect |
| the **evidence weight** | the raw evidence itself, log s | **held fixed at 1 throughout.** A single weight here would capture uniform over- and under-*inference*; splitting it in two would capture Griffin & Tversky's (1992) strength-versus-weight distinction. It is the natural next axis, not one this paper varies. §7 needs it by name to state what the **tempered-Bayes** family — the generalized-Bayes rules that raise the likelihood and the prior to powers and renormalize (§7) — can and cannot reach |

**An absolute score, and what a same-cue choice does to it.** The equation above is an *absolute* latent score — a judged strength for one cue–outcome pairing, natural for a rating. A two-outcome *choice* under one shared cue does not evaluate that score once; it evaluates it for each candidate outcome and compares. Write the score for the outcome that occurs as $J_1$ and for its complement as $J_0$ — swap P(O) for 1 – P(O) and s for s′, and keep the same cue, so P(C) is identical in both lines:

$$J_1 = \gamma[\log s + \rho_{pr}\log P(O) - \rho_{ev}\log P(C)], \qquad J_0 = \gamma[\log s' + \rho_{pr}\log(1 - P(O)) - \rho_{ev}\log P(C)]$$

A same-cue choice turns on the difference, and the cue-marginal term is identical in both lines, so it drops out of the subtraction:

$$J_1 - J_0 = \gamma\,[\log(s/s') + \rho_{pr}\,\mathrm{logit}\,P(O)]$$

$-\rho_{ev}$ log P(C) has cancelled *exactly, for every value of $\rho_{ev}$* — not only at a corner — and the only fact used is that both lines share one cue, and so one P(C). (The logit of P(O) is the log of its odds, log[P(O)/(1 − P(O))] — the outcome's base rate written on the scale a choice works in.) The proposition below generalizes this cancellation.

An expression that puts its *own*, separately-fitted weight on log s′ — $\gamma$[log s − $\lambda$ log s′ + $\rho_{pr}$ logit P(O)] with $\lambda$ a free parameter — is not another way of writing $J_1 - J_0$ above. The two coincide at exactly one setting, $\lambda = 1$, where the free weight is simply the full one the contrast already carries; everywhere else $J_1 - J_0$ minus the free-weight expression is $-\gamma(1 - \lambda)$ log s′, which is nonzero at every admissible false-alarm rate. Note what that difference does *not* contain: $\rho_{ev}$ has already cancelled from $J_1 - J_0$, so the agreement turns on $\lambda$ alone and holds for every value of $\rho_{ev}$ (closed form verified against 200,000 random configurations; the residual is zero to $7 \times 10^{-15}$ at $\lambda = 1$ and grows linearly in $1 - \lambda$ away from it). An asymmetric log s′ weight is therefore a genuinely different model from the absolute-score equation above, not a rewriting of it — which is exactly the odds-space parameterization set out below, and the reason this paper indexes its two weights by the space they are read out in rather than letting one name carry both.

The two readings do coincide at one setting, and it is worth seeing why. At the cue-corrected, prior-free corner ($\rho_{ev} = 1$, $\rho_{pr} = 0$) the absolute score becomes **lift**, log[P(O | C) / P(O)] — how far the cue raises the outcome above its base rate — and the same-cue contrast becomes $\gamma$[log s − log s′] = $\gamma$ log(s / s′), the signal-detection log-likelihood ratio. Only the first face belongs to $\rho_{ev} = 1$. Full cue correction is what turns sensitivity into validity for each outcome, so lift is a property of that corner. The second face is not: $\rho_{ev}$ has already cancelled from the contrast above, so at $\rho_{pr} = 0$ the contrast is $\gamma$ log(s/s′) for every value of $\rho_{ev}$. The signal-detection face therefore belongs to the whole prior-dropped column, and the weight that pins signal detection at 1 is the odds-space weight set out below, not $\rho_{ev}$. One corner, two faces — lift for the absolute score, detection for the choice — and only lift is specific to $\rho_{ev}$ = 1.

> **Proposition (same-cue choice invariance).** For any $K \geq 2$ candidate outcomes sharing one cue, normalized choice among them is invariant to the cue-marginal weight $\rho_{ev}$: the term $-\rho_{ev}$ log P(C) is common to every outcome's score under a shared cue and cancels from every pairwise contrast and from the full choice distribution, for any softmax temperature (the scale setting how deterministic the choice is) and any $K$. More generally it holds for any choice rule that depends on the scores only through their differences, softmax included. This generalizes the two-outcome result above; it is not restricted to binary choice. Same-cue choice data can identify the evidence and prior contrasts; it cannot identify $\rho_{ev}$. (Proved by the cancellation just stated; also verified numerically for $K$ up to 6 candidate outcomes and a range of softmax temperatures.)

**The kind of judgment a study collects determines which weight it can measure, and the two readouts measure different parameters.** A graded *rating* — "how likely is the outcome, given this cue" — expresses the whole equation, cue term and all, so it carries $\rho_{ev}$ directly. A two-choice *identification* — "given the cue, is it outcome A or outcome B" — is exactly the same-cue

contrast above, and by the proposition just proved the cue term cancels from it. A *different* coefficient survives in that contrast. Written in odds space, the decision variable is $\gamma[\log s - \rho_{ev}^{odds} \log s' + \rho_{pr}$ logit $P(O)]$, with the weight attached to the false-alarm likelihood $\log s'$ rather than to the cue marginal $\log P(C)$. These are two parameterizations of one inversion, not two spellings of one equation, so a choice identifies the odds-space weight $\rho_{ev}^{odds}$ and never the probability-space $\rho_{ev}$ of the equation above (the companion paper's Proposition B1 shows what the odds-space weight means: the tempered-Bayes family is exactly the surface on which $\rho_{ev}^{odds}$ equals the evidence weight, so $\rho_{ev}^{odds} = 1$ when that weight is 1, and a learner off that surface is discriminative rather than a re-tempered Bayesian, scoring outcomes directly instead of normalizing a tempered posterior).

**The two spaces meet at one corner, and that is not enough to carry a direction across.** At full correction they agree: cue weight 1 in the equation above is validity, $\log P(O \mid C)$, and weight 1 in odds space is the log posterior odds, which is validity written for a choice. That single coincidence is the one §2 has already used above. At the other end they do not agree. Setting the probability-space weight to 0 leaves the raw co-occurrence $\log P(C, O)$ (prior kept); setting the odds-space weight to 0 leaves $\log[P(C, O) / P(\neg O)]$, which exceeds it by $-\log[1 - P(O)]$ and can never equal it for any base rate strictly inside the unit interval. So the two scales share one endpoint, not two, and one point is not an interval along which a direction could transfer.

**The cancellation therefore costs more than the size.** Because the cue term is common to every candidate outcome under a shared cue, it drops out of every difference — so the choice probabilities are a *constant* function of the probability-space cue weight, and the Fisher information for that weight is exactly zero, for any number of alternatives and any softmax temperature. A same-cue two-choice test does not bound the weight loosely, or fix its sign while missing its magnitude. It carries nothing about it. Two learners at opposite ends of that scale — one applying no cue correction, one applying it in full — produce identical choices at every cell of every same-cue design.

**The design condition for a choice task.** Compare two cue–outcome pairings rather than two outcomes under one cue, and the cue term does not vanish: it becomes $-\rho_{ev} \log[P(C_1)/P(C_2)]$, a regressor the experimenter controls. **The choice depends on the probability-space cue weight if and only if the two alternatives' cue frequencies differ** (given a nonzero gain and a softmax choice rule at finite temperature) — necessary and sufficient, so a design with two distinct cues that happen to share a base rate is as blind as a same-cue design. If they differ, and the design also meets the rank condition of Appendix A.4a, the weight is identified by the ratio of two fitted coefficients, so a cross-cue choice recovers its *value* and not only its direction, with only the overall gain lost to the choice rule's temperature. Appendix A.4a gives the rank condition, a six-cell design meeting it, and the two ways an experimenter can meet the letter of the condition and still learn nothing.

**A scope condition: the cancellation applies to choice proportions.** It is a claim about the *choice*, not about everything a two-choice task records. Sequential-sampling models split a binary decision into a drift rate, a boundary and a starting point, and nothing here forbids the cue's frequency from surviving in the starting point after it has cancelled from the contrast. A starting-point effect of that kind would be a third route to $\rho_{ev}$ rather than a threat to the identity, and this paper does not develop it. The claim is the narrow one: the choice *proportion* does not depend on $\rho_{ev}$, so the weight cannot be recovered from choices alone (Figure 3).

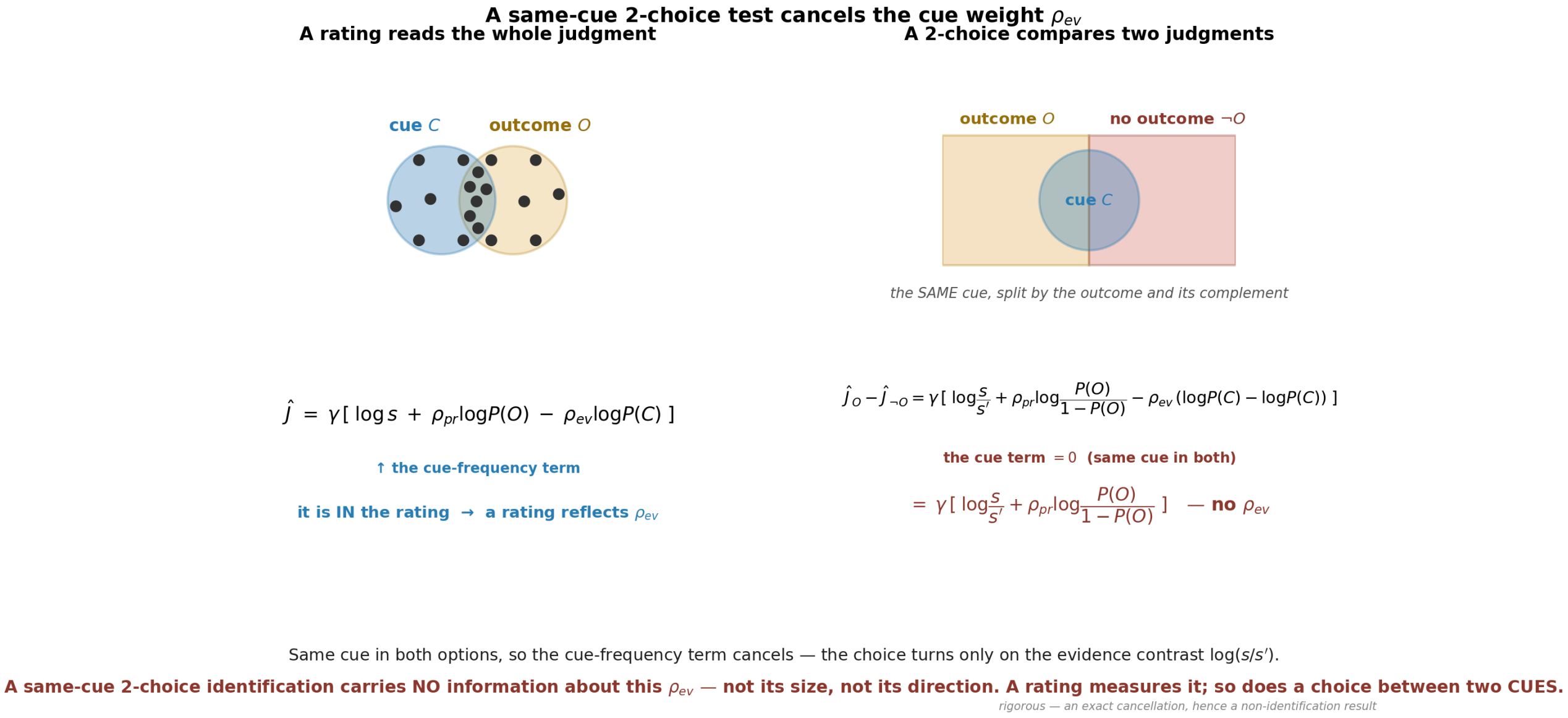

*Figure 3. A same-cue two-choice test cancels the cue weight; a rating retains it.*

**A second scope condition: the Proposition covers a comparison of outcome scores, not every binary response.** A same-cue choice depends on the scores only through $J_1 - J_0$, and that is the case the Proposition covers. A trial-by-trial yes/no prediction ("will the outcome follow this cue?") falls under it only if the learner forms it the same way, by weighing the outcome against its complement. If the learner instead compares the single score $J_1$ with a fixed criterion, nothing is subtracted, and the cue term does not cancel. It moves the operating point, the place on the evidence scale where answers switch from no to yes. Response format is known to matter for cue frequency. Perales, Catena, Shanks & González (2005) varied the cue's probability with contingency held fixed and found that the discriminability (d′) of trial-by-trial predictions did not reliably change ($p = .09$), while their response criterion and the numerical causal judgments did. That result concerns the prediction task as a whole. It does not settle whether same-cue predictions at zero contingency shift, which is the case the Proposition covers. It does show that a binary prediction can carry a cue-frequency effect, so the cancellation is stated here for a choice between outcomes and not for yes/no prediction in general.

**The scale is not the only thing that matters; the question is too.** Vadillo, Musca, Blanco & Matute (2011) held the response scale fixed — both measures were ratings, collected from the same people at the same stage — and found the cue-density bias *only* in outcome predictions: reliable there, absent in causal judgments, with a significant interaction between judgment type and cue density. Their design also makes the result hard to explain away. Varying cue frequency while holding contingency and the outcome's base rate fixed forces the validity P(O|C) to differ between groups — it was .59 where the cue was frequent against .81 where it was rare — and the authors state they chose

those values so the difference worked *against* a cue-density effect. The group with the lower validity judged higher anyway, which is under-correction for the cue's own frequency showing through a design built to suppress it. On the reading here the two questions are not asking for the same quantity. "How likely is the outcome, given the cue" asks for the validity itself, so the cue term enters at full strength; "how strongly does this cue cause the outcome" invites a contrast against the outcome's baseline. That contrast removes the prior term, not the cue term: with the outcome's base rate held fixed, as it was here, the equation above gives both questions the same latent cue-density effect unless the two draw different cue weights or gains. Two rating questions need not recover the same $\rho_{ev}$, and §6's design therefore fixes the question as well as the scale.

**Both weights need this point of care, because "correct" depends on the question, not just the dial.** If the question asks for a probability — "given the cue, how likely is the outcome" — the right prior weight is 1, and $\rho_{pr} < 1$ is neglect. If the question is prior-free — "how much does the cue *raise* the outcome, regardless of how common the outcome is" — the right prior weight is 0, and any $\rho_{pr} > 0$ is the prior leaking in where it should not: the outcome-density effect, read as an error against this prior-free target. So $\rho_{pr}$ is one dial read against two different targets, and classical base-rate neglect and the outcome-density effect are its two ends — the cue-density effect belongs to the other weight, not to this one. The cue weight is target-indexed the same way, just along the other axis of §4's table: $\rho_{ev} = 1$ is correct for both questions that fully correct for the cue (validity/posterior-probability *and* lift — the two differ only in $\rho_{pr}$), while $\rho_{ev} = 0$ is correct for both that do not (sensitivity and raw contiguity). So a single fitted $\rho_{ev}$ or $\rho_{pr}$ is neglect, intrusion, or exactly calibrated depending on which of the four corners the task is asking about — the table in §4 is the full statement of which pair is "correct" for which target, and every claim of "neglect" below should be read against it rather than against a single universal value.

## 3. The two effects, each on its own

The overlap between the cue and the outcome — how often they co-occur — can be divided two ways. Divide it by the cue and you get **validity** (of all the times the cue occurred, how often did the outcome follow). Divide it by the outcome and you get **sensitivity** (of all the times the outcome occurred, how often was the cue there). Same overlap on top, a different base rate underneath. In the case drawn, where the cue circle is the larger of the two, dividing by it returns the smaller number. That is the whole of the effect, drawn once, in Figure 4. The relative size of the two circles decides which ratio comes out larger, and Figure 5 varies exactly that.

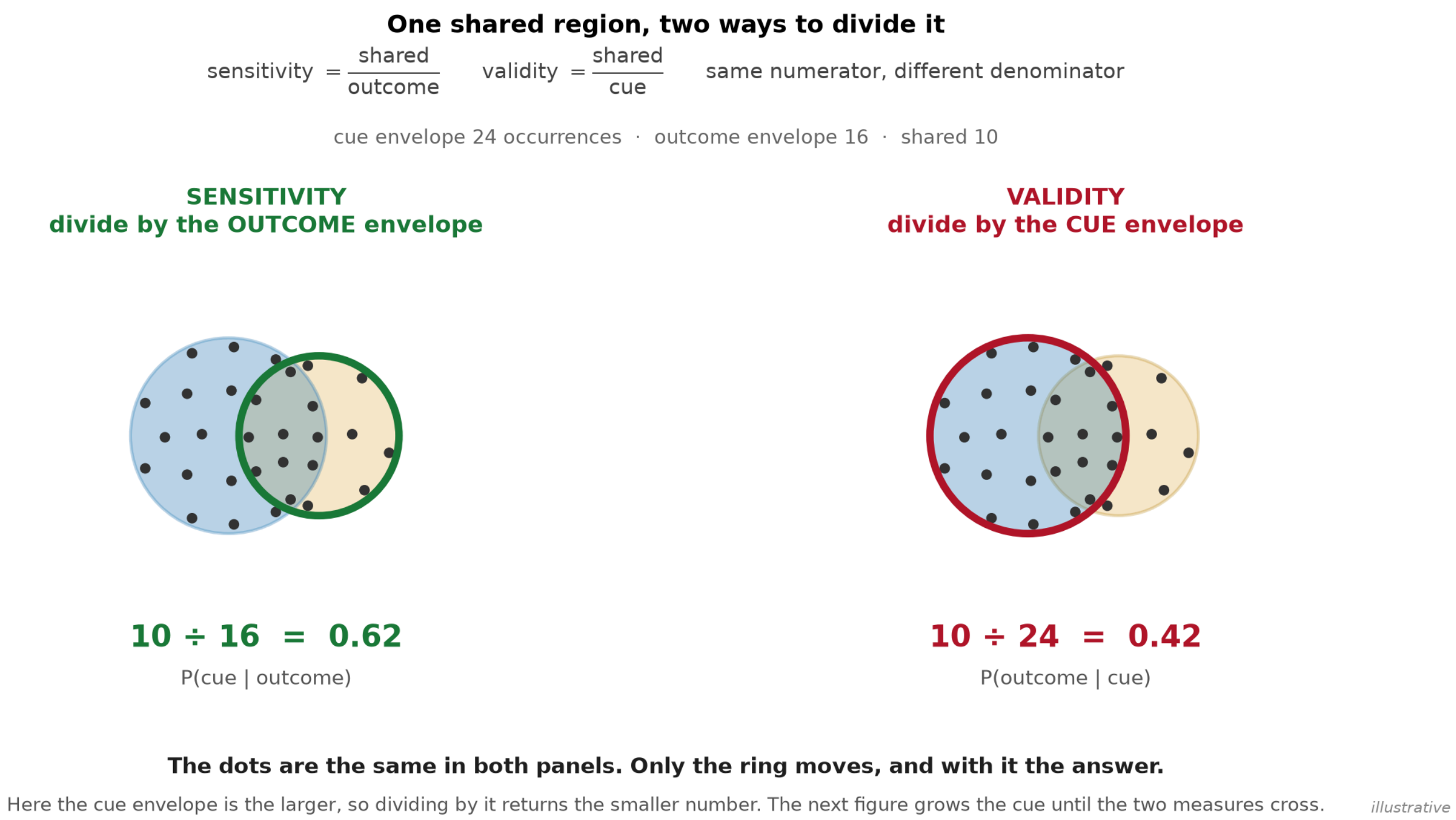

*Figure 4. The same overlap divided two ways: validity by the cue, sensitivity by the outcome.*

In the crossover picture the cue grows *and* more of it falls on-target, so the overlap grows with it. Validity falls, because the cue grows faster than the overlap it contains. Sensitivity rises, because the overlap grows while the outcome stays fixed. The two move in opposite directions because they divide by different base rates.

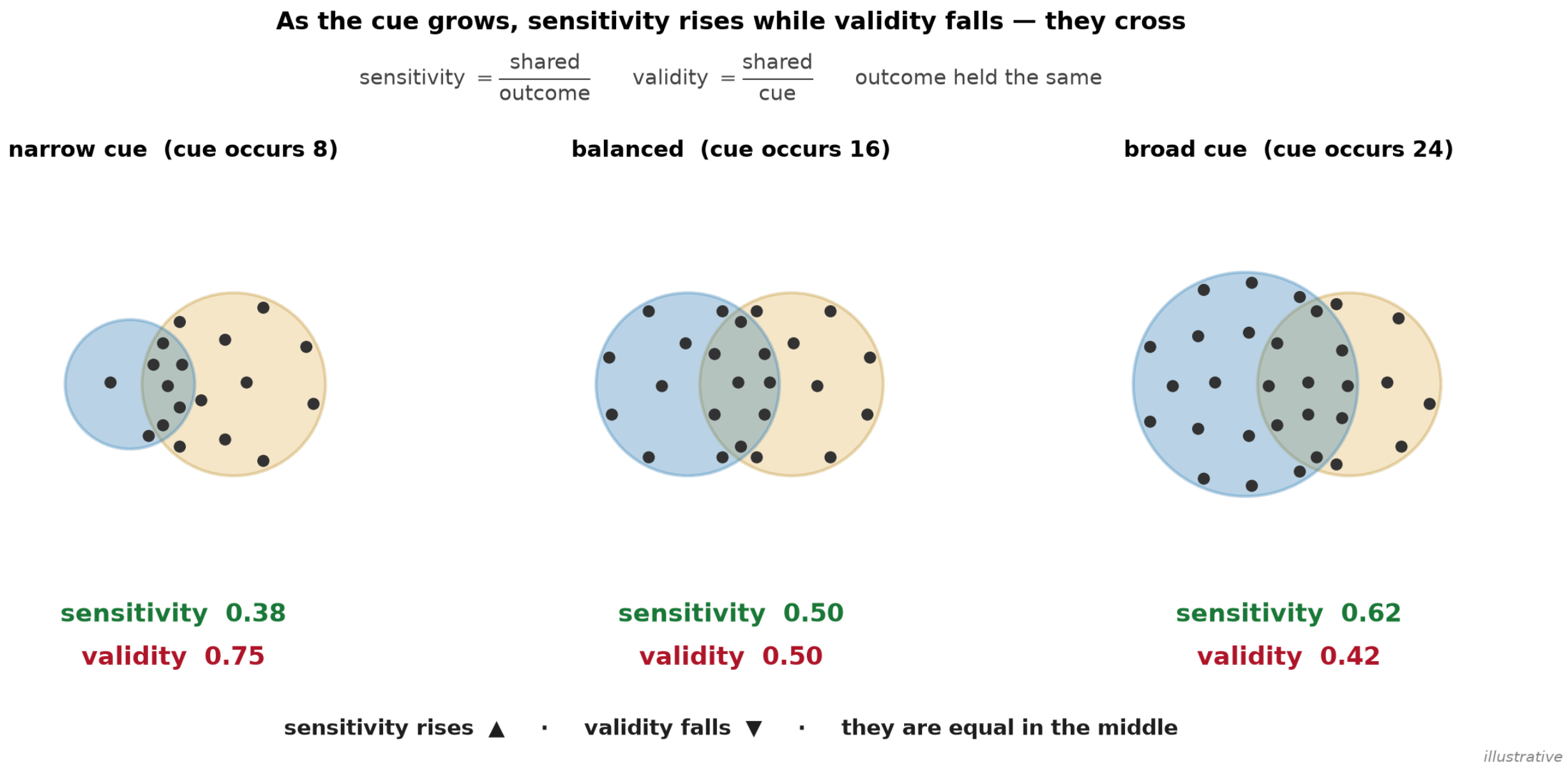

*Figure 5. As the cue grows, sensitivity rises while validity falls: they cross.*

This split, validity and sensitivity responding to different base rates, is the heart of the framework, and it comes in two versions that must not be run together. The first is just arithmetic: because validity divides by the cue and sensitivity divides by the outcome, growing the cue circle changes only the cue-divided ratio and growing the outcome circle only the outcome-divided one. That is true of any two normalizations of a shared overlap; it is about the *quantities*, not about people — which is all Figure 6 draws. The second, empirical version is the actual hypothesis: that the two *weights* a learner applies, $\rho_{ev}$ and $\rho_{pr}$, are genuinely distinct coefficients — that the cue-frequency regressor needs one of its own and the outcome-frequency regressor the other, rather than both terms sharing a single number. This is a claim about behaviour, not arithmetic, known here as a **two-coefficient separation**: the crossed design tests whether the data need two free weights or only one, $H_0$: $\rho_{ev} = \rho_{pr}$. A single-parameter account, one dial for "how much do people ignore base rates", predicts the two manipulations move one number together, which this test can reject. Coefficient separation is not yet the stronger, causal **double dissociation** of the neuropsychological sense — that would take interventions that make the cue marginal and the outcome prior separately salient (§7 names one candidate design) — but it is the framework's sharpest and most falsifiable near-term claim, and, as §6 makes explicit, it has not yet been tested.

One asymmetry has to be stated here rather than left to §6, because it qualifies what the two manipulations can be expected to do. The weights sit on separate terms. That separation is exact, and it is a statement about the *regressors* $\log P(C)$ and $\log q$, writing q for the outcome's base rate P(O) as the appendix does throughout. The experimenter, though, does not set those directly: the cue marginal is derived, $P(C) = s \cdot q + s' \cdot (1 - q)$, so $\partial P(C)/\partial q = s - s'$. Raising the outcome's frequency therefore drags the cue marginal along with it whenever the cue carries any contingency at all. The consequence is a clean split on one side only. A cue-side manipulation moves the $\rho_{ev}$ term alone; an outcome-side manipulation moves both, and the cross-term vanishes only at $s = s'$, which is null contingency — not the regime the identifying design runs in. The separation is thus triangular at the level of manipulations even though it is diagonal at the level of regressors, which is precisely why the association-strength cells of §6 are mandatory rather than a refinement.

The asymmetry carries a condition on how the experiment must be *analysed*, and the condition is strict. Fit the judgment without a term for the cue's own frequency, which is what an analysis that regresses a rating on the outcome's base rate alone does, and the recovered prior weight is not merely noisy but biased. Appendix A.2 gives the bias in closed form and its size on this paper's own cells: an observer who is *perfectly calibrated*, neglecting nothing, is recovered as a severe prior-neglecter. On a frequency-only design the bias is unbounded for observers near $\rho_{ev} = 0.5$ (Appendix A.2). The separation is therefore a claim about weights recovered from a model that carries the cue-marginal term on a design meeting the rank condition of Appendix A.4 — or from a design that removes the leak by construction, which Appendix A.3 shows how to build. Stated without either proviso, the claim fails generically once $\rho_{ev} \neq 0$: it survives only where the omitted cue-marginal term does not ride on the included regressors at all, or at an exact coincidence between those two projection coefficients and the prior weight itself — and that coincidence depends on the parameter the design is trying to recover, so no experimenter can arrange it (Appendix A.2).

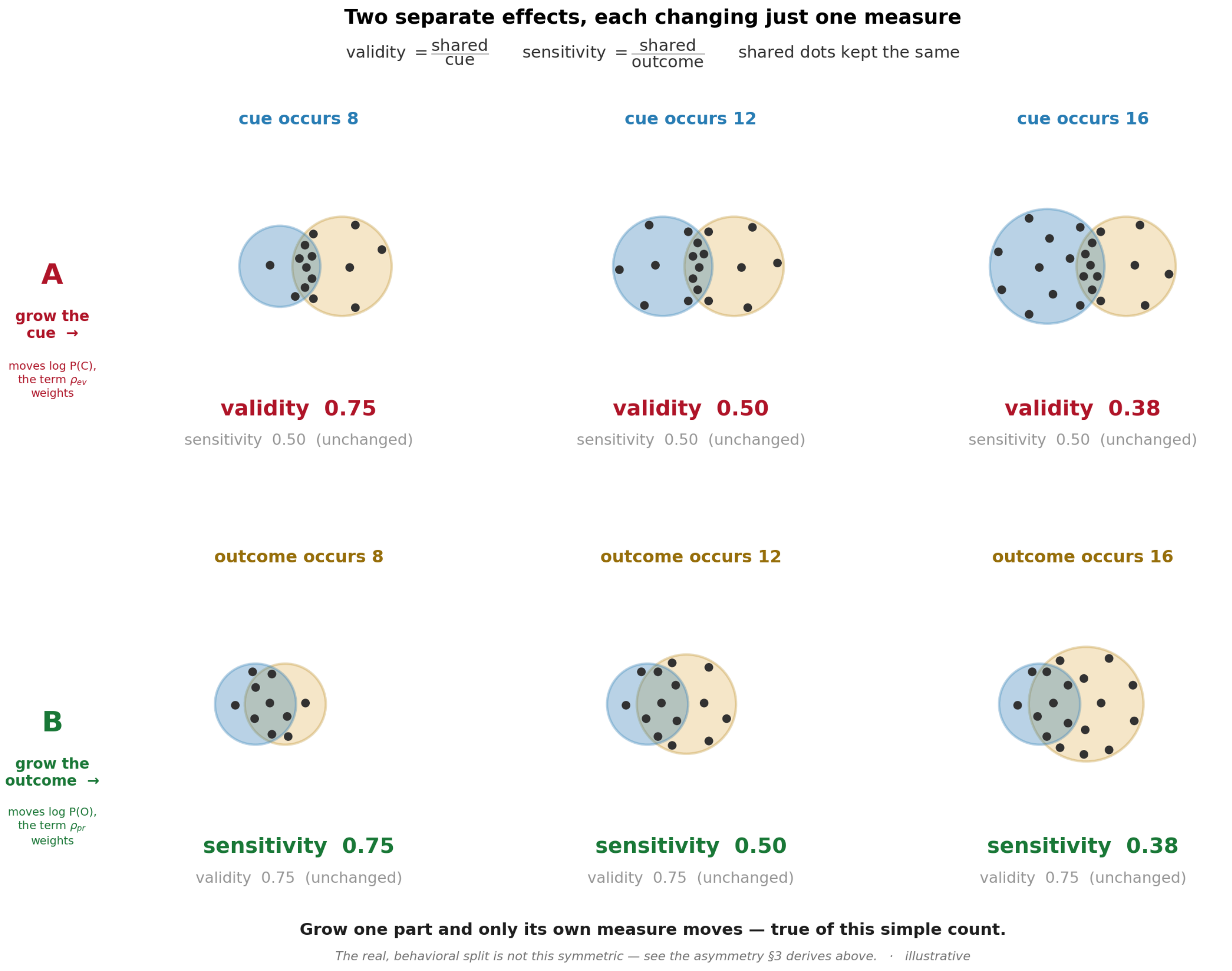

*Figure 6. Each marginal moves one measure alone — the arithmetic the two-coefficient separation is named for, not the asymmetric behavioural claim itself.*

Seen as behaviour, each weight has its own signature. Hold the evidence and the cue's own frequency fixed (the contour of Appendix A.3) and raise the outcome's base rate. On a prior-free question — how much does the cue *raise* the outcome — the correct judgment does not move, because neither the evidence nor the cue's frequency has changed; but a learner with $\rho_{pr} > 0$ is pulled upward toward the base rate — **prior intrusion** (Figure 7). (The target has to be named, per §2: against a probability question the right prior weight is 1 instead, and it is $\rho_{pr} < 1$ that is the error.)

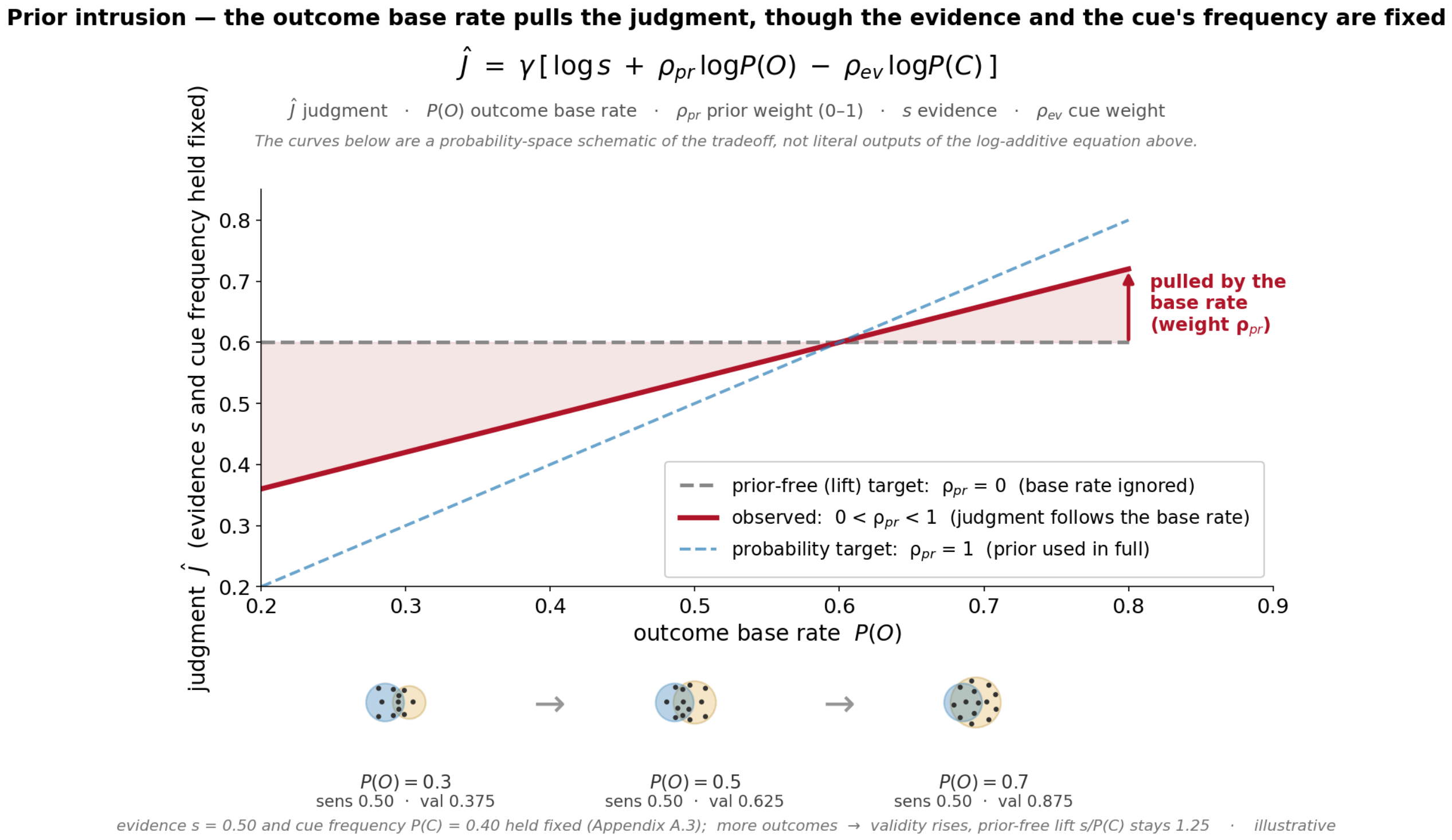

*Figure 7. Prior intrusion: the outcome base rate pulls the judgment.*

Now the mirror. Hold the evidence fixed and make the cue more common: the cue's true validity falls, the same overlap spread over a bigger cue, so the judgment should fall with it. A learner who under-divides keeps a common cue's strength propped up past what the current evidence supports — **cue intrusion** (Figure 8).

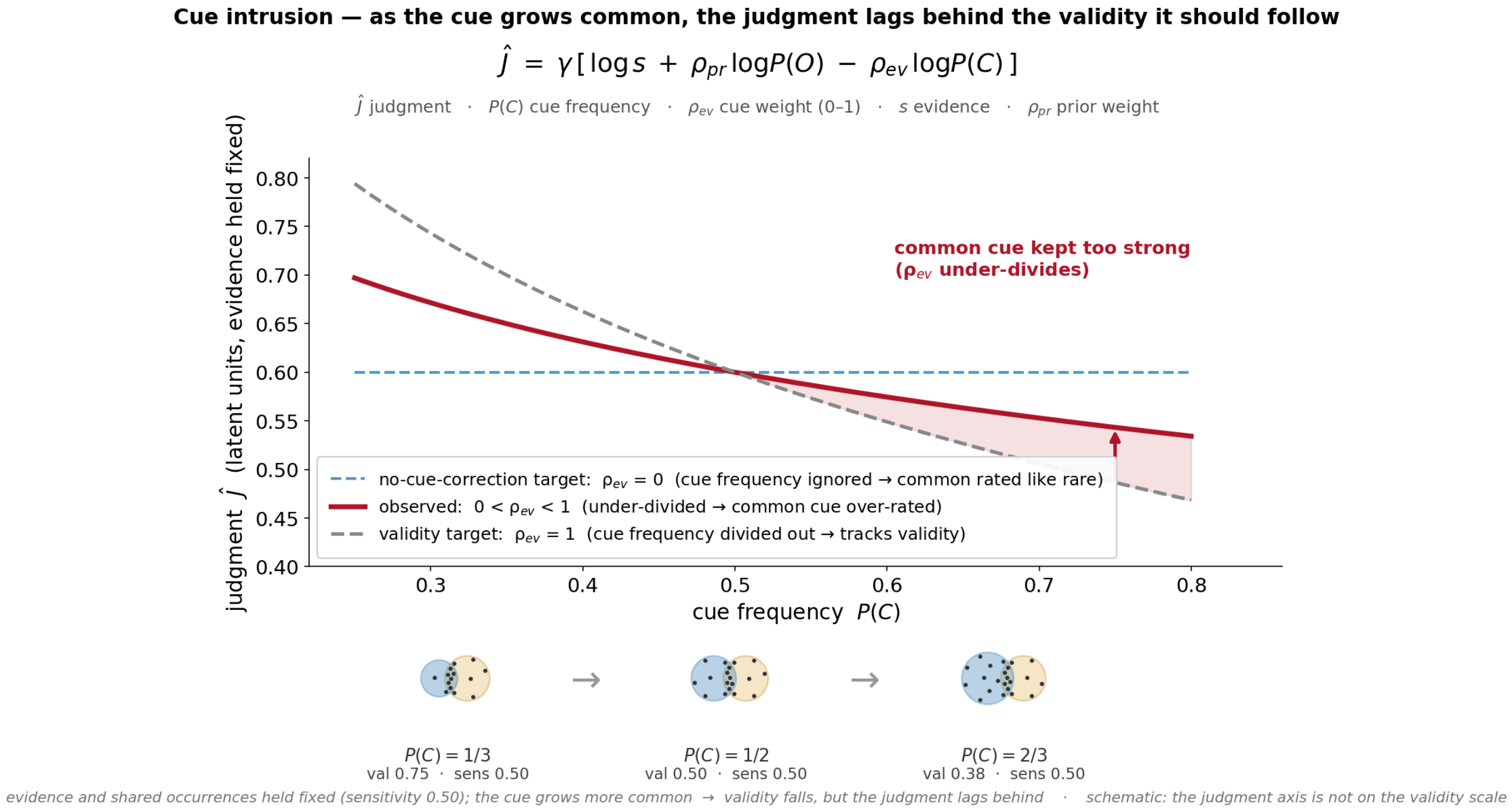

*Figure 8. Cue intrusion: a common cue's strength stays propped up past what the current evidence supports.*

The first is the outcome-density effect, the second the cue-density effect: the two weights, shown as behaviour.

## 4. The four corners

Setting each weight to an extreme, 0 or 1, gives four quantities the literatures already know, one per corner:

| | **prior kept ($\rho_{pr}$ = 1)** | **prior dropped ($\rho_{pr}$ = 0)** |
|---|---|---|
| **cue corrected** ($\rho_{ev}$ = 1) | **validity** P(O\|C) — the calibrated Bayesian | **lift** log[P(O\|C)/P(O)] — the rating's face of this corner |
| **cue not corrected** ($\rho_{ev}$ = 0) | **contiguity** — the raw joint, prior intact | **sensitivity** P(C\|O) — the raw evidence, log s, with neither correction applied |

In the prior-dropped column a same-cue choice reads the same thing in both rows: its contrast is the signal-detection log-likelihood ratio log(s/s′) for every $\rho_{ev}$ (§2), so what $\rho_{ev}$ = 1 adds there is lift, the absolute score's face. (The prior weight $\rho_{pr}$ is what controls the prior-odds contribution to the signal-detection *criterion* — in the same-cue contrast the offset is $\gamma \cdot \rho_{pr} \cdot$logit P(O), a product, so $\rho_{pr}$ is the dimensionless weight on that contribution rather than the criterion itself. It slides across the whole square, and $\rho_{pr}$ = 0 is the neutral setting for the criterion, adding no prior offset, not one of its extremes.) The framework is the *inside* of this square, not only its corners. Real learners sit somewhere in between, and the measurement of §6 recovers where: how far each weight falls short of 1. The corners also explain why one object wears four names across the three literatures: they are four readings of a single overlap, differing only in which base rate is divided out.

## 5. One coordinate across six accounts, and where it comes apart

§2 wrote the cue-frequency weight as a free parameter. It is not special to any one theory: the same weight can be found inside six standard accounts of learning and memory. The framework is therefore not a rival to those accounts but one shared quantity that can be located inside each of them (in signal detection, as its odds-space counterpart). The way each account contains the weight also shows where the accounts agree and where they do not.

| account | where the cue-frequency weight lives | how exact the match is |
|---|---|---|
| Bayes / information theory | the weight on how surprising the cue is on its own (its surprisal, –log P(C), large when the cue is rare) | exact — the weight *is* the parameter |
| signal detection | the full log-likelihood ratio log(s/s′) — how much better the evidence fits one answer than the other | an edge in odds space: $\rho_{ev}^{odds} = 1$. Its log(s/s′) holds at every probability-space $\rho_{ev}$ once the prior is dropped, so a rating weight below 1 does not count against it |
| Rescorla–Wagner (Rescorla & Wagner 1972) | how far learning has progressed — and the endpoint is not a bare number but a function of how often the outcome follows the cue | exact at the ends (early = none; asymptote = full when the two learning rates are equal); a schedule in between |
| memory-trace (MINERVA-2, a multiple-trace memory model; Hintzman 1984, 1986) | dividing the echo by its total intensity — a normalization we apply to MINERVA-2's own quantities, not one Hintzman performs | derived, and checked |
| prototype (Reed 1972; the abstraction effect itself: Posner & Keele 1968) | the prototype stores each category's average feature value, which tracks how often the cue occurs given the outcome; a category boundary corrected for the base rate gives full correction, an uncorrected boundary none | derived in the companion paper, and checked — Reed states the prototype rule (classify by distance to each category's prototype), not this reading of its boundary |
| exemplar (Medin & Schaffer 1978; generalized context model, GCM, Nosofsky 1986) | it already normalizes by construction: the ratio rule divides summed similarity to the category by summed similarity to all categories | sits at $\rho_{ev} = 1$ by default. Its selective-attention weights can move it below 1; whether those weights are the model's own or an addition to it is a question about the GCM, left open here, and no claim of this paper depends on the answer |

All six therefore map onto one axis. Among the accounts free to sit anywhere along that axis, an analysis that keeps only the coordinate cannot tell one from another — that is the sense in which the collapsed design does not choose. An account pinned at full correction would be a different case, because a weight recovered below 1 would already count against it. Figure 9 draws the two candidates for pinning apart from the spans for that reason, and neither is settled as pinned in this space: signal detection is pinned only in odds space, at $\rho_{ev}^{odds} = 1$, so a rating weight below 1 does not bear on it; whether the exemplar model is pinned turns on the open question about its attention weights.

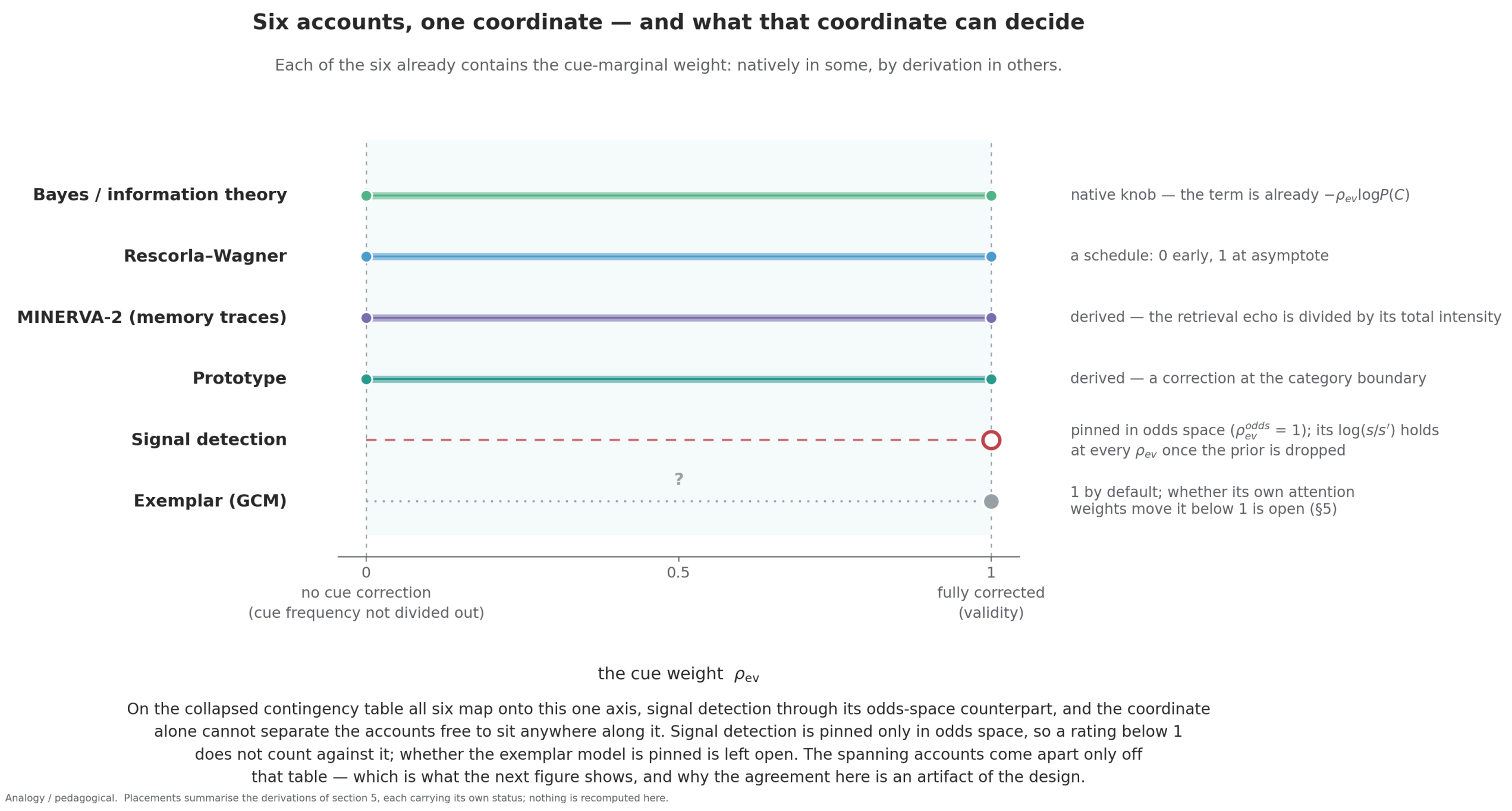

*Figure 9. Six accounts, one shared coordinate: that coordinate alone cannot separate the accounts free to sit anywhere along it. Signal detection is pinned only in odds space, and whether the exemplar model is pinned is left open.*

Two routes reach full correction: normalizing against the *competing outcome* (signal detection, the memory-trace echo, the exemplar sum) or dividing by the *cue's own frequency* (Bayes). Both land on validity, and you can convert one into the other — the same move from odds to probability used in §2. The accounts share how far a learner corrects, and differ in whether they read that out as odds or as a probability.

The table's Rescorla–Wagner row states an identity, in the model's symmetric case, not an analogy. Rescorla and Wagner give a cue's asymptotic associative strength as $\pi \beta_1 / [\pi \beta_1 + (1 - \pi) \beta_2]$, where $\pi$ is the probability of the outcome during the cue. The two learning rates are $\beta_1$ for reinforced trials, where the outcome arrives, and $\beta_2$ for nonreinforced trials, where it does not. Set them equal and the expression collapses to $\pi$ — the validity $P(O \mid C)$. The calibrated corner of this framework is that model's own equilibrium, in the model's own symmetric case.

But the accounts agree only on the surface. It holds only because the standard design collapses the data onto a 2×2 table, and on that table the cue-frequency weight is the *only* thing on which the accounts differ — so a four-cell summary cannot discriminate, through this scalar coordinate alone, the accounts that are free to sit anywhere along it. Those accounts may still disagree through their response map, learning trajectory, item memory, or other latent structure; what the collapsed table cannot do is show it. Give the design one more dimension and the accounts come apart. Two such dimensions are already available:

- **Learning over time.** An error-correcting (Rescorla–Wagner) learner starts by tracking raw co-occurrence and only converges to validity with training, so the difference between the high- and low-frequency conditions starts out pointing one way and reverses as learning continues (the sign flip of Figure 10, derived in the companion paper). The first half of that pattern, a density effect that shows up early, before learning settles, and fades with training, is shown for outcome density by Musca, Vadillo, Blanco & Matute (2010). An account that reads a fixed table never reverses.

- **Memory for specific items.** A learner that stores individual episodes (a trace or exemplar account) treats a repeated item differently from a novel one — a repeated-item advantage that grows with storage — whereas an account that keeps one strength per cue predicts no such difference (Figure 10).

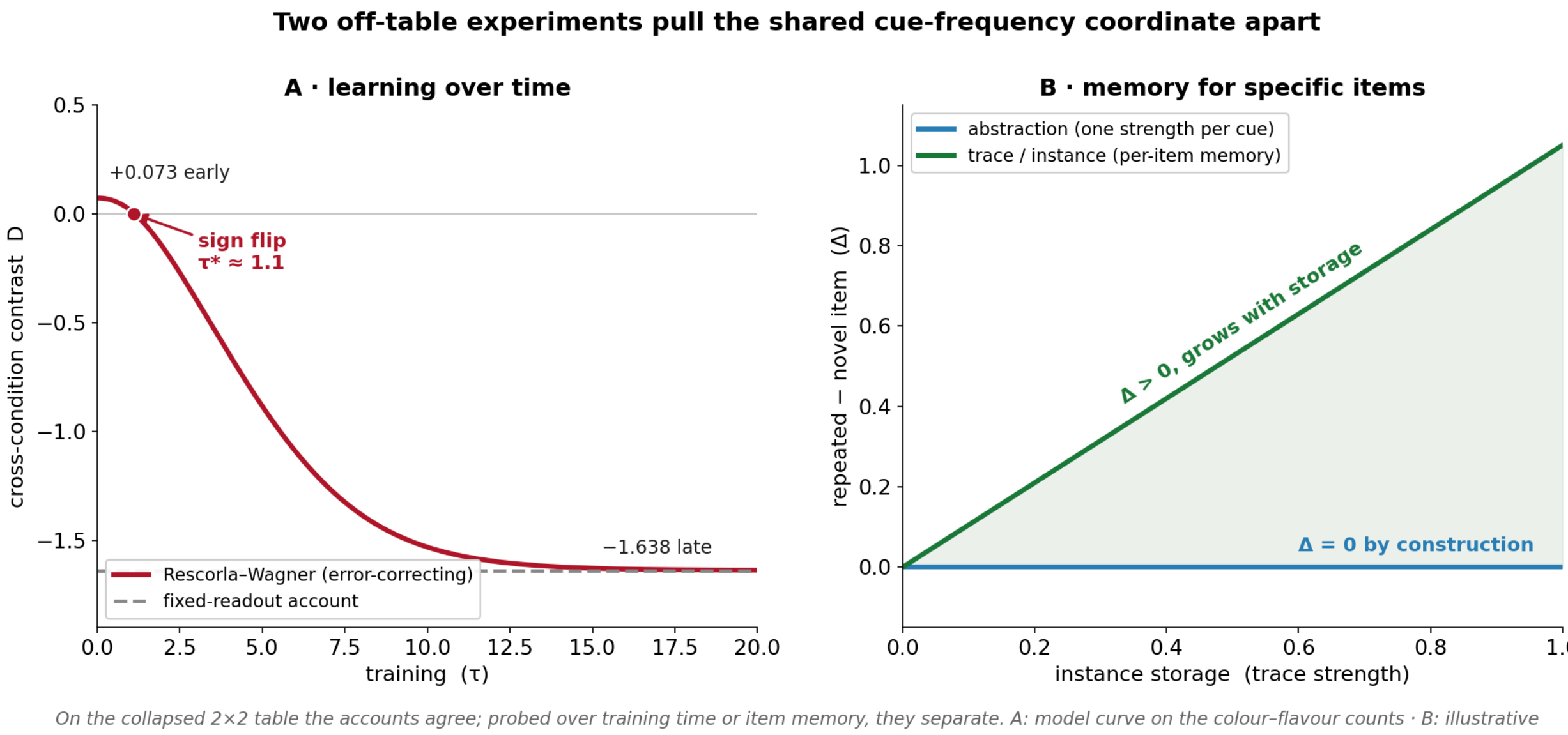

*Figure 10. Two tests separate the six accounts: learning sign-flip and item memory.*

So the shared weight is a coordinate, not a common cause. The two directions along which it stops being shared, learning over time and memory for specific items, are themselves experiments. (Full mappings and derivations: the companion paper's §3–§4 and its own appendix.)

## 6. Measuring both weights

The measurement follows from §2: because a same-cue two-choice task cancels the cue-frequency weight, estimating both weights together calls for graded ratings. Two other routes do carry cue-frequency information — a contrast between judgments made under *different* cues, where the term does not cancel, and process-level measures beneath the choice — but neither is developed here and neither is part of the design below. The design is a crossed one — cue frequency × outcome frequency × association strength — read out as ratings of how likely the outcome is, given each cue — the probability target of §2, for which the correct setting of both weights is 1. Concretely, nine core frequency cells cross three cue-skew levels, hit and false-alarm rates (s, s′) = (.83, .03), (.92, .08) and (.97, .17), with outcome frequencies .25, .50 and .75, and the cue's frequency follows from each; four association-strength cells add the weaker pairs (.75, .30) and (.70, .45) at outcome frequencies .25 and .75.

**All three factors are required, and strength is not optional.** Vary only the two frequencies and the design defeats itself. The cue's frequency, P(C), is not a free knob: it is fixed by how often the outcome occurs and how reliable the cue is, $P(C) = s \cdot P(O) + s' \cdot (1 - P(O))$. So raising the outcome frequency drags the cue frequency up with it. Across those cells the two base rates rise and fall almost together (they correlate about 0.93), which makes it nearly impossible to tell which one is driving the judgment — the variance-inflation factor, a measure of how badly two overlapping predictors blur each other's estimates, runs 49–55 across the nine core frequency cells. A factor of fifty means the sampling variance of the affected coefficient is about fifty times what an orthogonal design with the same residual noise would give it, so the problem is practical non-identification rather than a merely unattractive correlation. (The factor applies to the linear coefficients of that particular design matrix; it is not a universal uncertainty multiplier for the weights, which are ratios, and §A.5 shows the ratio intervals have to come from simulation rather than from a standard-error law.) (Those figures are properties of that particular design matrix, not of the model: a different arrangement of cells returns different numbers, so the design is named here and its generating code is deposited with the analysis. The 49–55 and 1.0 / 3.4 / 3.4 figures are the realized thirteen-cell design. The independent audit deposited with this paper builds its own abstract frequency-only grid instead, and returns 59.6 falling to 2.9 — different cells, the same collapse. A reader who runs that code should expect the second pair, not the first.)

In that regime the two weights cannot be told apart, and you cannot even be sure the cue weight is above zero. The association-strength cells change the actual cue–outcome contingency at fixed outcome frequency, which moves the evidence term off that line and breaks the tie: on the full thirteen-cell design the inflation factors fall to about 1.0 / 3.4 / 3.4, well-conditioned. A simulated single-point recovery check (synthetic data from a known truth, no participants) returns $\rho_{ev} \approx 0.22$ [0.09, 0.38] (clearly below 1) and $\rho_{pr} \approx 0.84$ [0.68, 1.07] from a true 0.20 / 0.70. That single interval still includes 1, but a full power sweep (Appendix A.5) shows why: it was an unlucky draw, not a design flaw.

**Both the recovery intervals just above and every sample-size number below come from a binary yes/no proxy rather than from the graded rating study proposed here; the collinearity figures are properties of the design matrix alone and do not depend on the readout.** The sweep simulates a yes/no response to the same design rather than the graded 0–100 rating, because the coarser readout runs in minutes; Appendix A.5 states the substitution and says the percentages should be recomputed under a bounded continuous or ordinal model before a sample size is fixed. Read as a planning aid at that order of magnitude: at the planning truth $\rho_{pr} = 0.70$ the planned thirty subjects give **90% power** to exclude prior calibration ($\rho_{pr} = 1$), and the estimator's mean bias stays within 0.03 of the true value across the sweep. The same sweep fixes the design's reach honestly: it separates *substantial* prior neglect from calibration, needs about forty-six subjects for a milder $\rho_{pr} = 0.80$, and **separates $\rho_{pr} = 0.90$ from calibration only at roughly 185**

**subjects, six times the planned sample**. The rating-model sweep is the outstanding task before the design is finalized.

Appendix A.4 states what the design must satisfy for the three parameters to be identified at all — a rank condition computable from the planned cells before a single subject is run, and therefore the pre-registrable form of the identifiability claim.

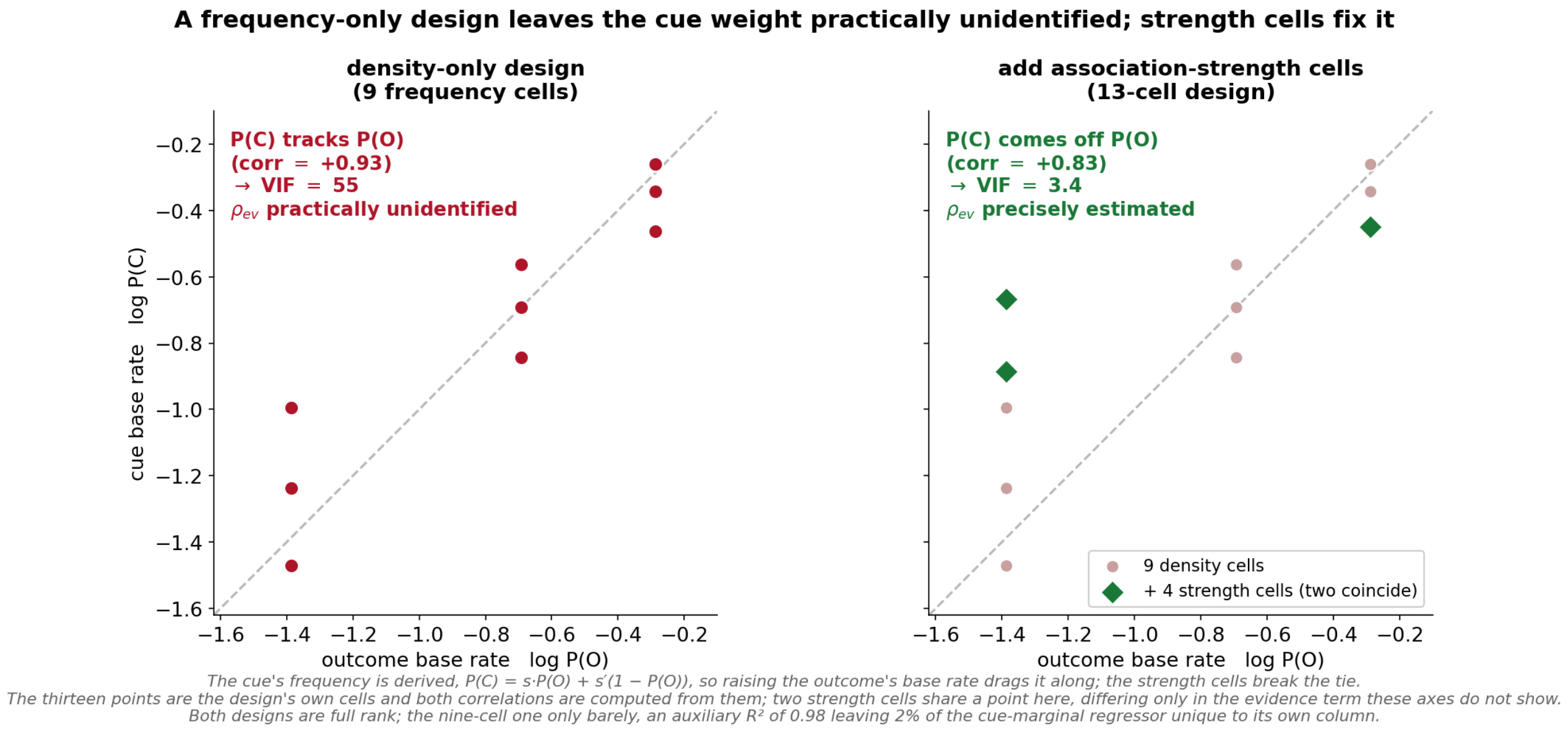

*Figure 11. A frequency-only design is full rank but leaves the cue weight practically unidentified; the association-strength cells make it precisely estimable.*

One point holds however the design is tweaked: the strength cells turn a setup where the two weights cannot be told apart into one where they can (Figure 11).

**The prediction, and how it can fail.** The framework's signature is the two-coefficient separation of §3, and it is worth being exact about what the experiment moves — including where the split is not clean. The *design* varies regressors: the cue-frequency manipulation moves log P(C) alone, but the outcome-frequency manipulation moves log P(O) and drags log P(C) along with it, by the asymmetry §3 derives. On this paper's own cells that drag is large, not a rounding effect: holding the cue's reliability fixed and raising the outcome's base rate from .25 to .75 moves P(C) by about .40 at every cue-skew level. Separating the two therefore rests on the association-strength cells, which move the evidence term off that line; that is what turns a collinear frequency-only grid into a design where both coefficients are estimable. The *coefficients* on those two terms are then estimated from the response surface, each holding the other regressor fixed.

The near-term test is whether one common correction coefficient is sufficient or whether the two marginal terms require distinct ones. (A later intervention targeting each normalization's salience could test the stronger claim that the coefficients themselves are selectively manipulable; that is a different experiment, and §3 keeps it separate.) Testing H0: $\rho_{ev} = \rho_{pr}$ is the clean nested version of that question (it will be preregistered; see Preregistration). Be exact about its reach: it tests the nested *common-weight* model inside this factorization, and rejecting it shows that one shared coefficient will not do here. It does not by itself rule out every one-parameter process model — an account with a different nonlinear mapping, unequal fixed loadings, or a different latent quantity can produce two apparent slopes without carrying two free weights, and §1.5's non-nested rivals are exactly such accounts. §7 states the prediction in that scoped form. This experiment has not yet been run: the prediction stands as a prediction.

There is also a cheaper test that needs no new experiment: whether a person's cue-density bias and their outcome-density bias correlate across people. Neither model fixes the sign of that correlation. Under the common-weight model the two zero-contingency slopes are $\gamma(1 - \rho)$ and $\gamma\rho$ (see below), so people who differ in the shared weight $\rho$ at a common gain produce a negative correlation and people who differ only in gain $\gamma$ a positive one; the sign alone separates nothing. Two separate weights do not predict its absence either, because both biases are scaled by the same gain $\gamma$, and people who differ in gain, or in numeracy, attention or measurement error, will show correlated biases even when their two weights are unrelated. Any dataset that already measures both biases in the same people — for example Shankar, Byrom, van Tilburg & Rakow (2025) — can check this, though at most as convergent evidence, not as a test that adjudicates the one- versus two-weight model by itself: a correlation of either sign fits the common-weight model, depending on whether people differ more in the shared weight or in gain, and a positive one also fits two weights that share a gain.

**That cheaper test is weaker than it looks, and the reason is worth stating.** The one-parameter account is not the only rival. A different family explains both density effects with no weight at all: judgment as probability theory corrupted by noise (Costello & Watts 2014), or as sampling with a regularising prior (Zhu, Sanborn & Chater 2020). Neither model was built for contingency learning — both target conservatism and the conjunction and disjunction fallacies — so the extension to the density effects is ours rather than theirs. It is a short extension, and the rival it names is real. A common noise source predicts that the two biases *positively correlate* across people — the same sign the common-weight model gives whenever people differ more in gain than in their shared weight. So the correlation bears only weakly on two weights against one, and does not separate either from noise. The crossed design does: a common noise source acting on a shared judgment gives no reason for the two marginal terms to need coefficients of different sizes. Selectivity across the two regressors, not the correlation, is what the framework stakes itself on.

That dataset already offers a second kind of support at the group level: fit to its zero-contingency density conditions, the two-weight model reproduces both density effects and their asymmetry — across its three experiments the fitted outcome-density slope runs about 2.2 to 2.5 times the cue-density slope — and within this parameterization a positive cue-density bias corresponds to $\rho_{ev} < 1$ under the stated readout and target assumptions, so the fit is evidence of cue-marginal under-correction in this zero-contingency regime (the fit is deposited with this paper, together with the checks below; the companion paper, Shavit (in preparation), develops it). Two things make that fit worth more than a single curve through four means. It **replicates**: the cue and outcome slopes reproduce across their independent samples and across two different designs, one manipulating both densities within subjects and one manipulating a single density. And it is **robust to position**: the cue weight comes out below 1 in both of the order-and-session splits that estimate it, weakest in the second-session cause condition, and in the collapsed cells and the independent sample besides. The outcome-over-cue asymmetry never inverts, its worst-case ratio still above 1. The gain $\gamma$ stays unidentified throughout, so these are bounds rather than point values — which is the identifiability discipline of Appendix A.4 showing up in real data.

**Zero contingency: what it can and cannot show.** At zero contingency $s = s' = P(C)$, so the judgment collapses to $\gamma[(1 - \rho_{ev}) \log P(C) + \rho_{pr} \log P(O)]$, and these cells identify only the two slopes on those logs, not $\rho_{ev}$ and $\rho_{pr}$ separately — $\gamma$ still has to come from elsewhere, per Appendix A.4. That limits what a one-versus-two-weight comparison can claim here: any pair of positive slopes fits the correctly specified common-weight null, $\rho_{ev} = \rho_{pr}$, exactly, by setting $\gamma$ to their sum and the shared weight to the outcome slope's share of that sum. So the zero-contingency slopes cannot reject one shared weight in favour of two free ones; a narrower comparison that instead forces the two slopes to be numerically *equal* is a stricter, different test — the special case where the shared weight sits at one-half, which is the weight setting of Hattori & Oaksford's (2007) dual-factor heuristic (§1.5) — not a test of the common-weight null. These cells *do* establish one thing cleanly, direction: a reliable positive cue-density slope forces $\rho_{ev} < 1$ whatever the size of $\gamma$ — the slope on log P(C) is $\gamma(1 - \rho_{ev})$, so its sign pins the weight's side of 1 while its magnitude stays confounded with the gain. The *sign* of $\gamma$ is doing work here, not just its non-vanishing: at $\gamma < 0$ the same positive slope would mean over-correction instead, which is why §2 fixes $\gamma > 0$ rather than $\gamma \neq 0$. This is the basis for calling the fit "evidence of cue-marginal under-correction in this zero-contingency regime" above — a statement inside this parameterization, not a re-description of the behavioural effect. Showing the two weights are genuinely distinct, rather than one shared weight, needs the full-rank rating design of this section, not the zero-contingency cells alone. Two further outcomes would break the model, and are named in advance rather than explained away later. First, *leakage*: if a condition-wise fit finds the outcome-frequency manipulation shifting both estimated coefficients while the cue-frequency manipulation shifts neither, the two are not separate dials but tangled together, and the claim that each correction is carried by a coefficient of its own fails. Second, *drift*: the two weights should stay put across the strength cells (the model holds the evidence term separate from them); if instead they re-fit to a new value at every level of contingency, they are not coordinates of neglect but fudge factors absorbing something the model got wrong. Weight-stability across the strength cells is therefore a stated adequacy criterion, not an afterthought.

**An open objection.** Blanco, Matute & Vadillo (2013) report that the two density effects combine *super-additively* — the bias from raising both frequencies exceeds the sum of the two alone — which looks like an interaction the additive model forbids. There is a standard reason it need not be: a rating is a bounded (0–100) readout of the additive judgment, and a bounded readout is curved, so below the scale midpoint two same-direction additive pushes produce a super-additive output and above it a sub-additive one — the interaction can live in the ruler rather than the weights (Figure 12).

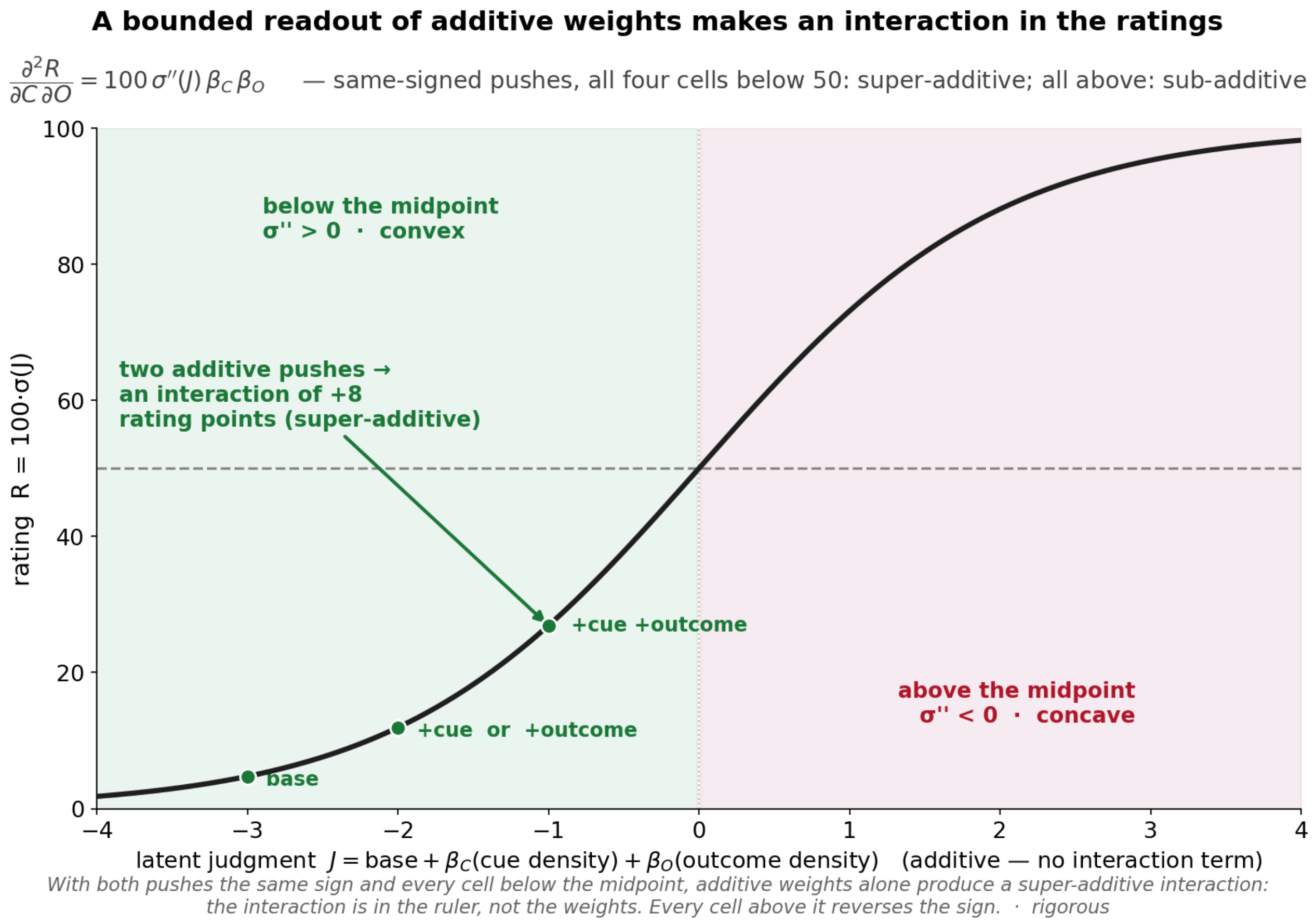

*Figure 12. Below the scale's midpoint, additive weights read as super-additive ratings: the interaction is in the ruler, not the weights. Above the midpoint the sign reverses.*

The bounded-readout curvature account does not, however, predict the interaction these particular cells show. Their cell means straddle the midpoint (Experiment 1, a 0–100 scale) or sit at and above it (Experiment 2, a −100 to +100 scale) rather than sitting below it, and the cells carrying the interaction fall on the side where the curvature works the other way. So the framework does not explain their result away. It commits to the bounded readout in advance and adjudicates on the fitted weights, and the sign of the interaction across the scale range stays a live test rather than a settled one. That sign is not a rule of thumb: Appendix A.7 gives it as an exact identity, so "does the cell rectangle sit on one side of the midpoint or straddle it" is a computable question, not an eyeball judgment.

The colour–flavour study (Shavit & Marks, in preparation), as reanalysed in the companion paper (Shavit, in preparation), checks only the odds-space weight: with the outcome base rate held flat, the cue tracks its own likelihood rather than the outcome's validity, which places the learner toward the un-corrected end **of the odds-space scale**. Because that study is a same-cue two-choice identification, §2's cancellation applies in full, and the probability-space weight — the weight the framework's equation carries — is not measured there at all, in size or in direction. The two scales share only their fully-corrected endpoint, so no reading transfers from one to the other at the un-corrected end where this result sits. That is exactly why the rating experiment above is worth running, and why a choice experiment contrasting two *cues* would serve as well.

## 7. Discussion

**The split's payoff, offered as a map to test rather than a result.** Read as a 2×2 of "which base rate is under-corrected," the four corners suggest a placement for phenomena that usually live in separate literatures: classical base-rate neglect may correspond to the outcome under-corrected; the cue-density effect, the illusion of control (Langer 1975), superstition (Skinner 1948), the illusion of causality behind pseudoscientific belief (Matute, Yarritu & Vadillo 2011) and the illusory correlation of clinical signs (Chapman & Chapman 1967) to the cue under-corrected; and the inverse fallacy of §1.4 (the mammography problem; Casscells, Schoenberger & Graboys 1978; Eddy 1982), the prosecutor's fallacy (Thompson & Schumann 1987), each of which reads a likelihood as a posterior, and distinctiveness-based illusory correlation (Hamilton & Gifford 1976) to both. **None of these placements is derived from the model here, and no dataset behind them is fit in this paper** — they are hypotheses for extension, and the coordinate's value is that it makes each of them testable rather than that it has already sorted them. The fourth corner is not an error at all: correcting for both base rates is Bayesian diagnosis done right, and natural-frequency formats often move people toward it on the classic problems (Gigerenzer & Hoffrage 1995), though the improvement is neither universal nor complete. A taxonomy of errors that cannot name its own success case is incomplete, and the grid gives that case its own corner (Figure 13).

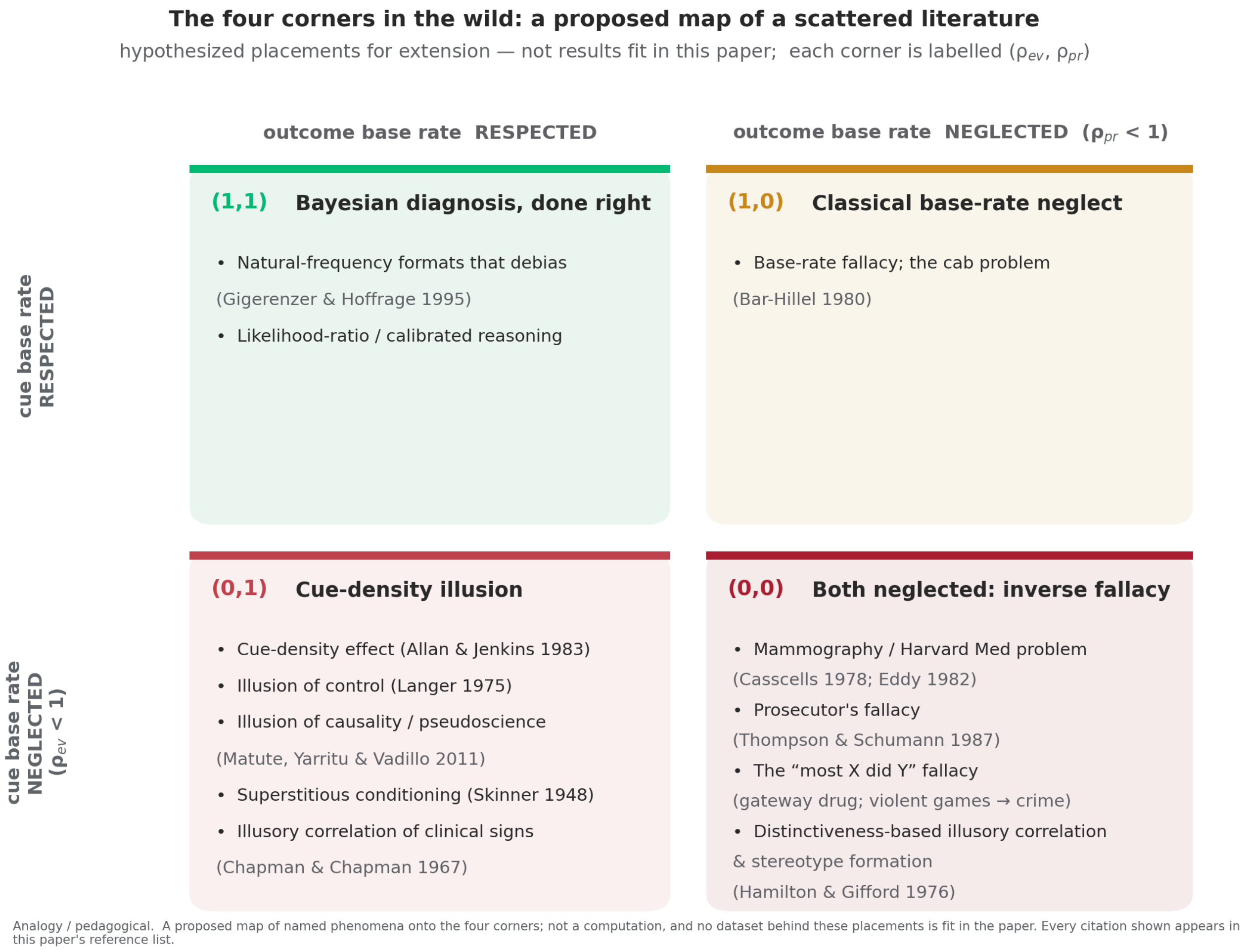

*Figure 13. A proposed interpretive map: where a scattered literature might sit on the 2x2. The placements are hypotheses for extension, not results fit in this paper. Each corner is labelled (cue weight, prior weight).*

The same picture also dissolves an apparent choice. The prior-free corner reads two ways, depending on the task. In a decision it is the signal-detection *decision variable* — the log-likelihood ratio log(s/s′) the chooser weighs, with the criterion offset at its neutral setting precisely because the prior has been dropped (§4); because $\rho_{ev}$ cancels from a same-cue choice, a decision reads log(s/s′) anywhere along the prior-dropped column, not only at this corner. In a rating it is *log lift*, or pointwise mutual information — how much the cue raises the outcome above its baseline. That is the corner exactly, and it is worth not blurring it: ΔP and Cheng's (1997) causal power are causal-strength measures, with different baselines and different normalizations, not approximations of it. They agree with log lift on the *sign* of a dependency, but not on its order. (Sign agreement is exact, not simulated. Since $P(O|C) - P(O) = [1 - P(C)] \cdot \Delta P$ — the cue lifts the outcome above its baseline by exactly ΔP, scaled down by how often the cue is absent — log lift and ΔP must share a sign whenever $P(C) < 1$. That condition is sufficient, and it is also necessary for the claim to carry content: at $P(C) = 1$ the difference is identically zero. Causal power shares the sign wherever its denominator $1 - P(O|\neg C)$ stays positive. Only the *order* claim rests on simulation. Drawing 200,000 cue–outcome configurations with both conditional probabilities uniform on [0.02, 0.98] and P(C) uniform on [0.05, 0.95], log lift and ΔP disagree about which of two cases is the stronger for about 12% of pairs, and log lift and causal power for about 17%. That rate moves with the sampling assumption — to about 6% and 9% under mid-range contingencies alone, to about 26% and 35% under positive contingencies alone — so it is an order of magnitude, not a constant. A strictly increasing transform of log lift disagrees on no pair in any sample, which is simply what monotone means; that zero is the baseline these percentages are read against, not a separate finding.)

So the framework need not decide whether people detect a signal or judge how far a cue raises an outcome: two-choice identification data live in one space, ratings in the other, and one inversion underlies both. None of these phenomena is new here. The density and pseudocontingency literatures found the two neglects; signal detection supplied the criterion. The new element is the split, writing the two neglects as two weights in one equation, and its signature prediction: that the common-weight constraint is inadequate under an identified design. The closest formal precedent is a signal-detection analysis of contingency assessment (Jozefowiez, Urcelay & Miller 2022), which finds that cue density shifts the *criterion* while associative interference and nonreinforcement lower *sensitivity* (d′, the signal-detection sense, not the hit rate s) to the contingency, under the decision variable the authors chose (see the caveat below); Allan, Siegel & Tangen (2005) found outcome density moving the criterion and contingency moving sensitivity, and predicted the same criterion locus for cue density. This is a dissociation of the same family. But a single SDT criterion absorbs *both* base-rate neglects, the cue's and the outcome's, into one scalar; it cannot tell an under-corrected cue frequency from an under-used prior. The factorization adds the separation itself: two independently-identifiable weights, one on the cue marginal and one on the prior. That separation turns the two-coefficient separation of §3 into a measurement. The contrast is cue-neglect against prior-neglect, then, not density against d′. Fiedler and colleagues identified the two base rates (how often the cue occurs, how often the outcome occurs); the signal-detection tradition identified the criterion; neither splits the criterion back into two separately-identifiable weights, which is the one thing the coefficient separation of §3 turns on.

**A caveat on criterion versus sensitivity.** Behaviour cannot settle whether cue frequency is a *criterion* effect or a *sensitivity* effect. Signal detection's two knobs are defined only relative to each other, so a shift in one can be rewritten as a shift in the other (Maia, Lefèvre & Jozefowiez 2018). For cue density in particular, "cue density moves the criterion" holds under one of two mathematically equivalent decision variables and "cue density moves the signal" under the other, and the choice between them cannot be made on empirical grounds (Jozefowiez, Urcelay & Miller 2022). Maia et al. (2018) do judge the cue-frequency criterion result of Perales et al. (2005) the better-grounded case. The framework does not rest on calling the cue term a criterion. It rests on the separation: two base-rate corrections written as two weights, which a single criterion folds into one. That is what the measurement of §6 recovers.

**A group-level weight is a population average, and the population may not be one group.** Stengård and colleagues (2022) tested base-rate reasoning over a broad problem space and fitted individual participants rather than the group. They report a bimodal split: one set of people almost entirely ignores the base rate, another almost entirely accounts for it, with the first better described by a linear-additive rule and the second by a Bayesian one. If that split holds for the two weights proposed here, a single fitted $\rho_{pr}$ midway between 0 and 1 would describe nobody. It would be the average of two groups, not the setting of one dial. Nothing in the framework requires a unimodal population, and the design of §6 fits subjects individually, so the question is answerable rather than fatal. But it changes what a group-level weight is evidence for, and it should be settled before either weight is read as a property of *the* learner.

It is worth being explicit about what the strongest version of that result would cost. If individuals sort cleanly into two groups at 0 and 1, then the weight is not a graded quantity that people hold at intermediate values; it is a mixture parameter, and what it measures is the proportion of people using each strategy. The framework would survive as a description of the two strategies and of what distinguishes them, and its corner identities would be untouched. It would lose the reading of an intermediate weight as one person's partial correction. The individual fits of §6 bear on these two readings, but the spread of fitted weights does not settle them alone: estimation error can make a mixture of endpoint strategies look unimodal, and a graded population can look bimodal. The comparison has to be between an endpoint-mixture model and a graded model, each carrying that error, and we adopt the graded reading only if the data favour it.

**Beyond neglect.** The two weights running from 0 to 1 are the *neglect* square; the fuller picture is a 3×3 that also includes over-correction — weights above 1 (Figure 14). The pattern Edwards (1968) called conservatism, revising too little away from the prior, reads here as over-use of the base rate; over-dividing by the cue is its cue-side mirror. Neither is a milder form of neglect. They sit on the *opposite* side of the calibrated point, since neglect is a weight below 1 and over-correction a weight above it. One wrinkle: over-using the prior and *under*-using the evidence produce the same behavioural signature, so conservatism reads either as $\rho_{pr} > 1$ here or as a weight below 1 on the evidence term — a third knob this framework deliberately holds fixed. The two readings are not distinguished by any existing data; the companion paper names the diagnosticity manipulation that would separate them.

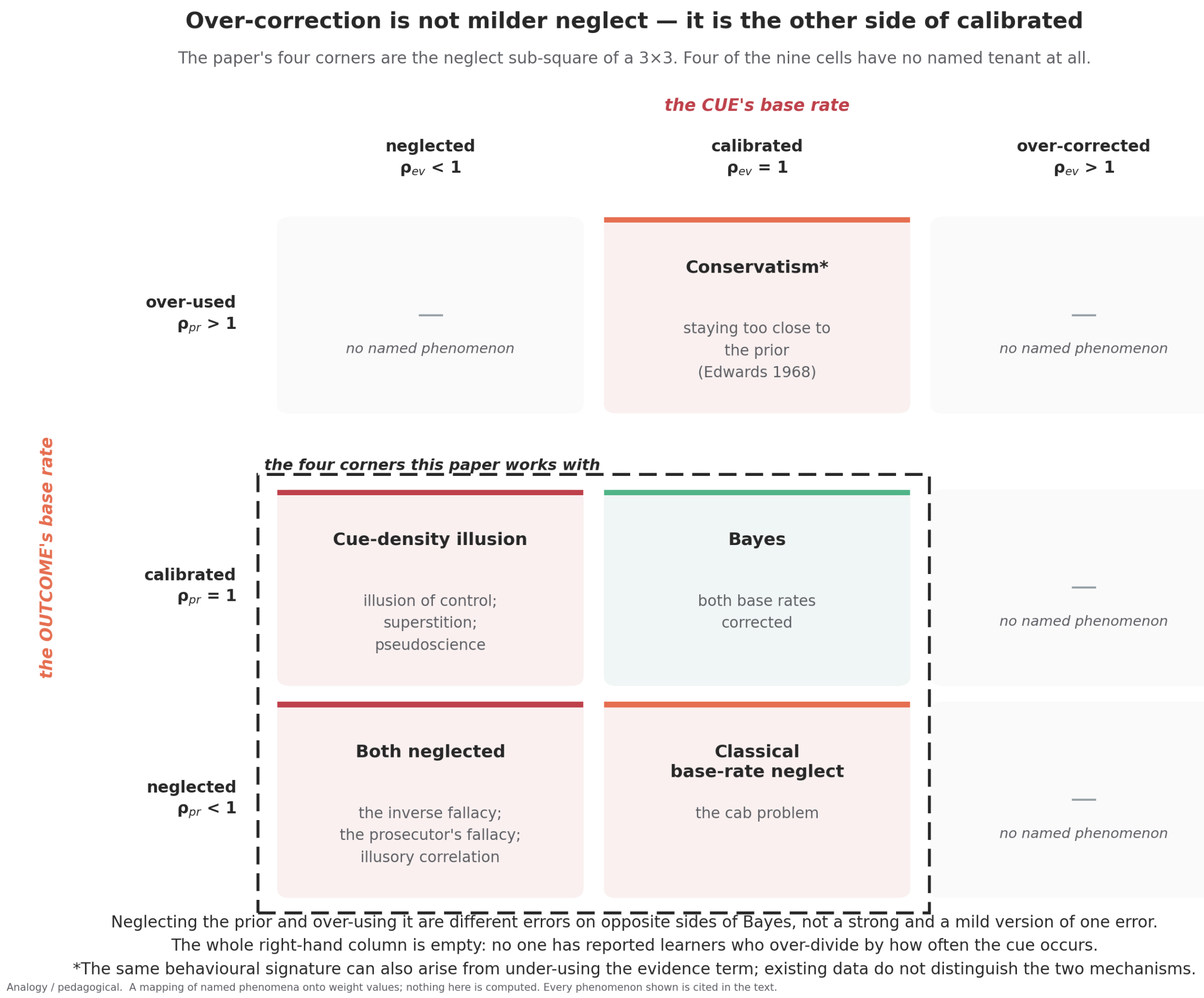

*Figure 14. The neglect square inside the 3x3: over-correction is the other side of calibrated.*

And the equation deliberately holds fixed one thing other two-parameter accounts vary: the evidence term itself. A third weight, on log s, would capture uniform over- and under-*inference*; splitting it in two would capture the strength/weight distinction of Griffin & Tversky (1992) — a natural next axis once the two base-rate weights are pinned. The framework is thus one face of a larger space; the claim here is bounded to the neglect corners and the coordinate they share, not a theory of every misjudgment.

**Two of the three weights are still Bayes; one is not.** These knobs are not the same kind of knob, and the difference is the sharpest form of the split. Turning down the prior weight, or re-scaling the evidence weight, is a *tempering* — raising the prior or the likelihood to a power, exactly the generalized-Bayes move (Grether 1980) — and it leaves the learner computing a genuine, if distorted, posterior, *provided the normalizer is allowed to follow*. That proviso is not a technicality. A normalized tempered posterior divides by a new partition function (normalizing sum), $\Sigma_j \; P(C \mid O_j)^a \; P(O_j)^b$, which is neither the original P(C) nor a fixed multiple of it; leave the old cue marginal in place and the scores no longer sum to one across outcomes, so the absolute-score form is not a posterior at all. On the choice or odds scale the new normalizer is common to every outcome and cancels, which is why the tempered family collapses there and why the claim holds in that space — the space the next sentence works in.

Turning down the cue weight does not. Because the cue's frequency $P(C) = \Sigma\, P(C \mid O)\, P(O)$ is a *derived* quantity, the total summed over outcomes, rather than an ingredient the learner multiplies in, no re-weighting of the prior and the likelihood can give it an independent knob: on the two-choice scale the whole tempered family collapses onto the single line where the *odds-space* cue weight $\rho_{ev}^{odds}$ equals the evidence weight (proved in the companion paper's appendix; the proof is also deposited with this paper, in the replication package, so a reader of the preprint can check it without waiting for the companion).

The space has to be named, because §2's cancellation means the probability-space $\rho_{ev}$ is not on the two-choice scale at all to be compared with anything. On that scale the statement is direct: the evidence weight is pinned at 1, so the tempered line is $\rho_{ev}^{odds} = 1$, and a choice rule with $\rho_{ev}^{odds} < 1$ sits off it. The shared endpoint cannot carry that verdict across to the probability-space weight, because one shared point transfers no direction (§2): a same-cue choice computed from the equation's own scores lands exactly on the tempered line whatever $\rho_{ev}$ is. The probability-space argument runs on the absolute scores instead. With $\rho_{ev} < 1$ they are not a normalized posterior, and no re-tempering of the likelihood and the prior reproduces them, because a tempered posterior's normalizer follows its exponents. So a learner with $\rho_{ev} < 1$ is not running a tempered Bayes rule at all, but a *discriminative* one whose denominator is only partly divided out — which is why it is the one axis the standard two-parameter accounts, evidence-weight and prior-weight alike, structurally cannot reach, and why only a rating, not a same-cue choice, can see it.

**Three levels, kept apart.** The claims here come at three levels, and the paper commits to only two of them. At the level of *mathematics*, Bayes' rule contains two logarithmic corrections, one per base rate. That is arithmetic, and it is not in dispute. At the level of a *behavioural model*, a learner applies each correction partially, and how much of each is the quantity §6 estimates. At the level of *cognitive implementation*, the framework makes no claim. Two fitted weights are equally consistent with two processes, one process with two outputs, a heuristic that happens to have this profile, or an associative rule that produces it as a by-product. The proposed experiment can test the behavioural parameterization — whether one coefficient or two are needed — and cannot reach the cognitive implementation beneath it. The third level stays open, and nothing here should be read as settling it.

**The additive form is an assumption, and the design should test it rather than presume it.** The two weights sit on separate terms of a sum. That is a claim about functional form, not a finding, and every result here inherits it. If a learner combines the two corrections in some genuinely interactive way, the weights this paper recovers are not two psychological quantities but the best additive projection of a process that is not additive. The two-coefficient separation would then be a property of the fitted model rather than of the person.

Two things keep that from being an article of faith. The first is that one familiar source of apparent interaction has already been separated out. A bounded rating scale curves, so additive inputs read as super-additive below its midpoint and sub-additive above it, which is why the interaction can sit in the ruler rather than in the weights. If the cells fall on the side where curvature predicts the opposite sign, the curvature account fails, and the interaction cannot be attributed to the ruler. The second is that the crossed design of §6 can adjudicate the form itself rather than assume it: fit the additive model and a model carrying an explicit cue-by-outcome interaction term to the same data, report both, and let the comparison decide. An interaction that survives on the side of the scale where curvature predicts the opposite sign is evidence against the additive form, and the framework should be read as refuted in that respect rather than rescued. We state the commitment here so that the test is not optional.

**The claim, plainly.** The psychology, that cue frequency and outcome frequency each distort judgment, belongs to the cue-density and pseudocontingency literatures; the criterion machinery belongs to signal detection. Neither is claimed here. The claim is the split, and its signature prediction is coefficient separation, not coefficient movement: across a design that independently excites the cue-marginal and outcome-marginal regressors, a common-weight model constrains the two corrections to one coefficient while the two-weight model allows them to differ. Coefficient separation is the one thing the framework stakes its correctness on, and it is what makes it falsifiable where a single absorbed criterion and a base-rate-matching heuristic are not.

---

## Appendix A — Estimating the weights: leakage, bias, design, and power

*The body promises that the two weights can be estimated apart. This appendix states the conditions under which that promise holds. Each result was derived and then checked numerically by an independent routine that shares no code with the model, and each is labelled by how strongly it is established. The generating scripts are deposited with the analysis.*

*A.1–A.4 are algebraic rather than simulated, and are not revisited below, but they do not all carry the same scope and the differences matter. A.1 is a property of the design's marginals alone and holds under any readout whatever. A.3 is an algebraic construction on those same marginals, and likewise. A.2 is a statement about a linear projection on the modelled latent or rating scale: under a nonlinear link the analogous pseudo-true coefficients depend on that link and on the outcome distribution, so the formula below is not the bias a misspecified nonlinear fit would show. A.4 gives structural identification for the stated linear predictor, and extends to a known injective response link under correct specification, because such a link can be inverted at the mean — not to an unspecified readout. A.5–A.6 are numerical power claims for one specific readout choice; see the note at the head of each for what that choice does and does not cover.*

*Notation, carried over from §2 and compressed here because these results are read as algebra: $q$ is the outcome's base rate, the $P(O)$ of §2's equation; $s$ and $s'$ are the cue's hit and false-alarm rates, $P(\text{cue} \mid \text{outcome})$ and $P(\text{cue} \mid \text{no outcome})$; $P(C) = s \cdot q + s'(1 - q)$ is the cue's own frequency, derived*

*from the other three rather than set; $\gamma > 0$ is the overall scale factor, $\rho_{pr}$ the prior weight and $\rho_{ev}$ the cue weight.*

## A.1 The leak from an outcome manipulation has a closed form

The experimenter does not set the cue's own frequency directly. It is fixed by the other two, P(C) = s·q + s′(1 − q), so raising the outcome's frequency q drags the cue's frequency up with it. The question is by how much. **Proved:** the elasticity — the percentage change in P(C) per percentage change in q — is exactly

$$\varepsilon \equiv d \log P(C) / d \log q = q(s - s')/P(C) = 1 - s'/P(C),$$

so the design-level response of the judgment to an outcome manipulation is $\gamma[\rho_{pr} - \rho_{ev} \cdot \varepsilon]$. The elasticity ε quantifies the leak. It sits in (0, 1) for a positively predictive cue; it is exactly zero at null contingency, s = s′; it approaches 1 as the cue becomes perfectly specific, s′ → 0; and it reverses sign for a negatively predictive cue.

**On this paper's own cells the leak is first-order, not a small correction.** Across the thirteen operating points ε runs from 0.12 to 0.95, and across the nine frequency-only cells from 0.54 to 0.95. An outcome manipulation there moves the cue-weighted term about as much as it moves the prior-weighted one. *(Admissible wherever the logarithms are defined; no s > s′ restriction.)*

## A.2 Omitting the cue term biases the recovered prior weight, sometimes without bound

Suppose the analyst fits the judgment on the outcome's base rate and the hit rate, but never models the cue's frequency — which is exactly what a conventional analysis of a rating task does. **Proved (rating readout):** the recovered prior weight does not converge to the truth but to

$$\hat{\rho}_{pr} \to ( \rho_{pr} - \rho_{ev} \cdot d_q ) / ( 1 - \rho_{ev} \cdot d_s ),$$

where $d_s$ and $d_q$ are the coefficients from regressing the omitted term log P(C) on the ones that were kept. They are properties of the design, not of the observer: they measure how strongly the omitted regressor moves with the included ones. (The expression presumes $1 - \rho_{ev} \cdot d_s \neq 0$; where that denominator crosses zero the limit is undefined, which is the unbounded case two paragraphs below.)

Setting that limit equal to the truth and clearing the denominator leaves $\rho_{ev}(d_q - \rho_{pr} \cdot d_s) = 0$, so for a *given* observer the bias vanishes when $\rho_{ev} = 0$ or when the two projection coefficients happen to satisfy $d_q = \rho_{pr} \cdot d_s$. Orthogonality, $d_s = d_q = 0$, is the stronger condition: it is what guarantees unbiasedness for *every* $\rho_{pr}$ at once, which is why it, and not the observer-specific coincidence, is the design target A.3 constructs. The coincidence is not a design one can build, because it depends on the very parameter the design is trying to recover.

Three consequences follow.

**An observer who neglects nothing is recorded as neglecting a great deal.** On the thirteen-cell design a perfectly calibrated observer ($\rho_{ev} = \rho_{pr} = 1$) is recovered as $\hat{\rho}_{pr} \approx 0.33$; on the nine core cells the recovered value goes negative. The apparent neglect is manufactured by the omission.

**On a frequency-only design the bias is unbounded.** There $d_s \approx 1.99$, so the denominator crosses zero near $\rho_{ev} \approx 0.5$ and the estimate passes through $\pm\infty$. An unbounded bias is not a small distortion to be noted and set aside.

**Both weights are exposed.** Omitting the outcome term instead biases the cue weight — a true pair (0.2, 0.7) is recovered as $\hat{\rho}_{ev} \approx -0.51$. The triangularity of §3 is asymmetric in *manipulations*; omission bias is not.

*Verified numerically two independent ways: the closed form against a directly computed projection (agreement 1e-10), and against simulated noisy ratings, whose mis-specified fit converges to the formula's value rather than to the truth as trials per cell grow from 50 to 50,000. Under a binary logistic readout the limit has no clean closed form, but at 20,000 trials per cell the calibrated observer is recovered as 0.334 against the rating-readout prediction 0.329 — same direction, same size, so in this check the bias belongs to the omission and not to the link.*

**Conditions under which the promised test is valid.** With the full three-term model on a design meeting A.4's rank condition, both least squares and the logistic maximum-likelihood estimator are consistent for all three parameters: the triangular design-level Jacobian then costs *variance* only, never bias. Carrying the cue-marginal term on a full-rank design, or removing the leak outright, is therefore **sufficient** for a valid separation test, and it is the only route an experimenter can guarantee in advance. It is not **necessary**: the bias also vanishes at $\rho_{ev} = 0$, at the observer-specific coincidence above, and — as the next paragraph catalogues — in whole classes of analysis that have no cue-marginal term to omit in the first place. Sufficiency is what a design can be built on; necessity is not claimed, and would be false.

**Existing measurements: what this does and does not say.** The result is a warning about a procedure, not an indictment of a literature. We checked, and the established estimates escape — each for a reason worth naming, because the reasons mark the boundary of the exposure. Analyses conducted on the *odds* scale are immune outright: the cue's frequency cancels from a ratio, so no term is left to omit, and the classical analyses that model choice through latent log posterior odds, linear in the log prior odds and the log likelihood ratio, are of this kind (Grether 1980). So is Griffin and Tversky's (1992) regression of the logarithm of judged log odds on the logarithms of sample strength and sample size, which is odds-scale for the same reason although its two regressors are different ones — how extreme the sample is and how large it is, rather than the prior odds. And so is the odds-scale reconstruction Benjamin, Bodoh-Creed & Rabin (2019) build from Griffin and Tversky's own published condition medians. The density studies cited here are immune for a different reason: they were run at null contingency, which is exactly where the leak ε of A.1 is zero (Blanco, Matute & Vadillo 2013; Musca, Vadillo, Blanco & Matute 2010; Shankar, Byrom, van Tilburg & Rakow 2025). And an analysis that weights the hit and false-alarm rates alongside the base rate omits nothing at all, since the cue's frequency is a function of exactly those three — the individual-participant modelling of Stengård, Juslin, Hahn & van den Berg (2022) is of this kind, and further benchmarks each weight against the best linear approximation to Bayes rather than against a nominal 1.

The result is therefore a condition on future work, including our own. The exposure is specific: a graded magnitude readout, a fitted weight on the outcome's base rate, a cue frequency that is derived rather than controlled, and a non-null contingency. That combination is precisely what §6 asks for — so the estimate this paper calls for is itself of the exposed kind, and is trustworthy only with the cue-marginal term in the model. We state the condition here rather than leave it to be discovered.

### A.3 A design that removes the leak by construction

**Proved.** Hold the cue's frequency fixed while the outcome's frequency moves, by solving $P(C) = \bar{P}$ for the false-alarm rate:

```
s′(q) = ( P̄ − s·q ) / ( 1 − q ),   with s held fixed.
```

Along this contour, varying q changes log q and nothing else in the judgment: log s is pinned and log P(C) is constant by construction. The slope of the judgment on log q is then $\gamma \cdot \rho_{pr}$ exactly, free of the cue weight, **even in a fit that never models P(C)** — the within-contour slope carries no cue-weight term by construction (on contour-only cells log s is constant and aliases the intercept, so the separate A.2 projection coefficients are not defined there). The cue-side dual is already clean: varying s′ at fixed s and q moves only the cue term, delivering $\gamma \cdot \rho_{ev}$. The contour is feasible with a positively predictive cue if and only if $s \cdot q < \bar{P} < s$; at $s = 0.70$ and $\bar{P} = 0.50$, outcome frequencies from 0.20 to 0.65 give false-alarm rates from 0.45 to 0.13, all admissible.

This turns the two-coefficient separation from a regression-adjusted claim into a raw manipulation-level one: two clean slopes deliver $\gamma\rho_{pr}$ and $\gamma\rho_{ev}$, hence the ratio of the weights with no adjustment at all. It does not replace A.4 — separating the weights from the overall gain γ still needs the hit rate varied somewhere in the design. Note also the trade: because s′ moves along the contour, the decision-form evidence log(s/s′) is not constant across its cells. The rating-form judgment is unaffected, since s′ enters it only through P(C), which is why the result is exact for ratings.

### A.4 Conditions for identifying the three parameters

**Proved.** Write the cell means as $\mu = \alpha + \gamma[\log s + \rho_{pr} \log q - \rho_{ev} \log P(C)]$ with a free intercept α. Then γ, $\rho_{pr}$ and $\rho_{ev}$ are jointly identified **if and only if the design matrix [1, log s, log q, log P(C)] has rank 4, and γ ≠ 0.** Identification needs only that the gain not vanish; the stronger convention $\gamma > 0$ adopted in §2 does separate work, licensing the one-sided readings of §3 and §6 rather than this rank condition, which is indifferent to the sign. Because P(C) is derived rather than chosen, the ways to fail are structural, and each is a design the literature actually runs:

| Failure | The design | What survives |
|---|---|---|
| Null contingency everywhere ($s = s'$) | $\log P(C)$ collapses onto $\log s$ | $\gamma(1 - \rho_{ev})$ and $\gamma\rho_{pr}$, so the weights only as the ratio $\rho_{pr}/(1 - \rho_{ev})$ |
| Perfectly specific cue ($s' = 0$) | $\log P(C) = \log s + \log q$ | $\gamma(1 - \rho_{ev})$ and $\gamma(\rho_{pr} - \rho_{ev})$ |
| Outcome frequency never varied | $\log q$ is constant | **the prior weight is lost** |
| Hit rate never varied | $\log s$ is constant | the gain $\gamma$ is lost; only $\rho_{pr}/\rho_{ev}$ |

The third row is the design the colour–flavour dataset uses, and the rank result turns an observation into a theorem: in a *rating* readout, holding the outcome frequency flat is exactly what removes the prior regressor and leaves the cue-side coefficient isolated. **The theorem is about the rating form, and it does not transfer to that dataset's actual readout.** The colour–flavour study collects a same-cue two-choice identification, and §2's cancellation removes the cue term from that contrast entirely, so flat outcome frequency cannot make $\rho_{ev}$ estimable there — the coefficient is simply absent from what a choice measures. The choice data identify instead the odds-space weight of §2. The two applications must be kept apart: flat outcome frequency isolates the cue-side coefficient *in a rating design meeting the remaining rank conditions*, and says nothing about a choice design. The second row carries a warning — a highly specific cue sits near an identifiability cliff, degenerate exactly at $s' = 0$ and numerically unusable just beside it.

**For the experimenter, positively stated:** vary the hit rate, vary the outcome frequency, and place cells off both the $s' = s$ and $s' = 0$ families. The thirteen-cell design does all four, and it is well conditioned: after standardizing the three non-intercept regressors, its **condition number** — the largest singular value divided by the smallest, the standard scale-free measure of how near a design sits to rank deficiency — is $\kappa = 3.4$ (a different statistic from §6's variance-inflation factors of 3.4, which share the value by coincidence), against $\kappa = 14.7$ for the nine frequency-only cells. The condition number is reported here rather than a raw smallest singular value, because a raw singular value carries no scale-free meaning. It moves with the units of the columns, and — less obviously — with which matrix it is taken from. On the design matrix that carries an intercept column and leaves the three regressors uncentred, these same thirteen cells return 0.41 in natural logarithms and 0.18 in base-10 logarithms, and 0.09 with the marginals written as percentages rather than proportions. Centre those columns first and the percentage figure comes back at 0.41 rather than 0.09, because writing a marginal as a percentage adds a constant to its logarithm and centring deletes constants. The two condition numbers just quoted come from a third matrix again, the standardized intercept-free one, which is why each is named where it is used. The raw value also grows with the square root of the number of cells, so designs of different size cannot be compared by it at all. The nine frequency-only cells are technically full rank but sit beside a combined deficiency: there an auxiliary regression of $\log P(C)$ on $\{1, \log s, \log q\}$ returns $R^2 = 0.98$, so barely 2% of the cue-marginal regressor's variance is unique to that column, against $R^2 = 0.70$ on the thirteen-cell design. That is the collinearity of §6 seen as geometry. Because rank and conditioning are both computable from the planned cells before any data exist, this is the pre-registrable form of the identifiability claim. A.8 carries this further, from a condition a design must clear to a statement of which design is best and what the cheapest working one gives up.

*Each deficient family was confirmed by singular value decomposition, and its surviving combination confirmed by constructing two different parameter triples that share that combination and produce cell means agreeing to within $9 \times 10^{-16}$ — floating-point precision, so numerically the same prediction — against a third triple that does not. The condition numbers, the auxiliary $R^2$, and the scale- and size-dependence of the raw singular value are computed by the same deposited script.*

## A.4a The condition for a choice task to see the cue weight

**Proved.** §2 showed that a same-cue two-choice test cancels the cue-frequency term exactly. The same algebra says what it would take to keep it. Let two alternatives be whole cue–outcome pairings, pairing *i* carrying hit rate $s_i$, false-alarm rate $s'_i$ and outcome base rate $q_i$, so that its cue frequency is $P(C_i) = s_i q_i + s'_i(1 - q_i)$. The latent contrast is

$$\gamma[\ \log(s_1/s_2) + \rho_{pr} \log(q_1/q_2) - \rho_{ev} \log(P(C_1)/P(C_2))\ ].$$

In words: comparing two pairings leaves three difference terms, and the cue weight now sits on the log ratio of the two cue frequencies. The condition follows and it is sharp:

**The choice distribution depends on $\rho_{ev}$ if and only if $\log P(C_1) \neq \log P(C_2)$, given $\gamma \neq 0$ and a softmax choice rule at finite temperature.** The condition is necessary as well as sufficient, so two distinct cues that happen to share a base rate leave the design as blind as a same-cue design does.

If it holds, and if the design matrix $[1, \Delta\log s, \Delta\log q, -\Delta\log P(C)]$ has rank 4, the two weights are recovered as ratios of fitted coefficients: $\rho_{pr} = \beta_q/\beta_h$ and $\rho_{ev} = \beta_c/\beta_h$, where $\beta_h$ is the coefficient on the hit-rate contrast. **A choice task therefore recovers the cue weight's value, not only its direction** — the overall gain $\gamma$ is the only casualty, confounded with the choice rule's temperature, which is the exact mirror of A.4's "hit rate never varied" row where the gain is lost and the ratio survives. The calibration test $\rho_{ev} = 1$ is then the linear contrast $\beta_h - \beta_c = 0$, matching the rating design's form.

**For the experimenter, positively stated:** contrast two cues, separate their frequencies as far as the apparatus allows, and vary hit rate, outcome base rate and false-alarm rate. A six-cell design drawn from feasible pairings reaches standardized condition number $\kappa = 1.11$, against $\kappa = 3.4$ for the thirteen-cell rating design of A.4 — better conditioned on fewer cells.

**Two designs the literature runs meet the letter of the condition yet identify the cue weight poorly or not at all.** First, if each physical cue appears at exactly one frequency and the model allows that cue its own additive bias, then $\log P(C)$ is a function of cue identity and the bias absorbs it completely: rank falls to 4 of 7. The remedy is to run the same physical cue under two schedules differing in hit rate, base rate **and** false-alarm rate, which makes the cue-frequency regressor vary *within* cue; that restores full rank at $\kappa = 5.21$, so robustness to cue-specific bias costs

about a factor of five in conditioning. (A.4's own theorem assumes a single free intercept and carries the same exposure; see the note below.) Second, if the two cue frequencies differ only slightly, the design remains full rank and **the condition number does not detect the problem**: across a two-hundred-fold shrink of the frequency separation, κ moves only from 6.05 to 6.82 while the asymptotic variance of the cue-weight estimate rises from 4.7 to $8.9 \times 10^5$. A condition number is a scale-free diagnostic for *rank*, and this is a loss of *information* rather than of rank. Pre-register the asymptotic variance of the ratio estimate, or the raw smallest singular value, alongside κ — not κ alone.

*Every claim in this subsection was verified symbolically, the necessity by quantifier elimination rather than by example, and every numerical design figure was computed from cells checked in exact rational arithmetic against $0 < s' < s < 1$. Each check carried a negative control that fired: an equal-frequency design falls to rank 3 with the cue regressor taking one value; two parameter triples differing in $\rho_{ev}$ by 0.55 alias to $1.1 \times 10^{-16}$ on that design and separate by 1.08 on the six-cell design; and the same-cue design's cue-weight column is identically zero. The script is deposited with the analysis, named in the deposit's own index, and it carries a self-test: it re-runs its three load-bearing checks on a design that violates the condition and requires each one to fail, so a reader can see that the checks can fail.*

### A.5 Power: what the design can and cannot detect

**A proxy, not the proposed instrument.** §6 proposes a graded 0–100 rating. The sweep below instead simulates a binary (yes/no) response to the same design — a coarser readout, chosen so the sweep runs in minutes. That substitution is not free: it trades away the continuous rating for a single-bit response, and the two need not have the same power. Treat the table as an order-of-magnitude planning aid, not a power guarantee for the rating study as proposed. A defensible version for the actual 0–100 measure would fit a bounded continuous or ordinal model (beta, logit-normal, or ordered-logit) to the planned number of ratings per participant, not to training trials, and the percentages below should be recomputed against that model before they are used to fix a sample size.

**Verified numerically, for the binary-readout proxy just described** (200 simulated experiments per point; thirteen-cell design, 45 trials per cell, binary readout, between-subject gain heterogeneity, and interval method as in the design check). Power is the probability that the upper end of the 95% interval falls below 1 — that the prior weight is shown to be genuinely below calibration.

| True $\rho_{pr}$ | N = 15 | N = 30 | N = 60 | N = 120 | N for 80% |
|---|---|---|---|---|---|
| 0.60 | 0.92 | 1.00 | 1.00 | 1.00 | ~11 |
| 0.70 | 0.67 | **0.90** | 1.00 | 1.00 | ~21 |
| 0.80 | 0.35 | 0.65 | 0.88 | 1.00 | ~46 |
| 0.90 | 0.14 | 0.20 | 0.33 | 0.62 | ~185 |
| 1.00 (calibrated) | 0.02 | 0.05 | 0.04 | 0.04 | — |

Four things follow, all **conditional on the binary-readout proxy above and to be re-checked against the actual rating model before the design is finalized.** The planned thirty subjects give **90% power at the planning truth of 0.70** under that proxy, so the single straddling interval reported in §6 was a tail draw rather than a design failure. There is a **detectability boundary near 0.8**: the design distinguishes substantial prior neglect from calibration, and separates 0.90 from calibration only at roughly 185 subjects, six times the planned sample — the planned sample licenses only the former. The false-exclusion rate comes out at 2–5% against a nominal 2.5%, which is **consistent with** the nominal rate but does not pin it down: at 200 replicates per point the exact binomial interval on the worst realized value, 10 of 200, is [2.4%, 9.0%], so this sweep can neither confirm nor rule out a modestly anti-conservative interval, and a calibration claim would need on the order of a thousand replicates. And the estimator's mean bias stays within 0.03 of the true value across the sweep, so any systematic tilt is small next to the sampling noise in the recovery reported in §6 — again, all under the proxy readout.

One methodological caution: normal theory applied to the delta-method standard error predicts *more* power than the simulation delivers, because the interval on a ratio of coefficients is right-skewed. Power claims for ratio-valued weights should come from simulation or a Fieller-type interval — an interval built for a ratio of two uncertain coefficients, which stays valid when the denominator is itself noisy — not from the standard-error law. That caution is about a weight reported as a ratio. A.8 shows the primary common-weight hypothesis need not be tested as one at all, because it is a linear contrast in the coefficients the design already estimates.

### A.6 The same question asked of the cue weight

The sweep above tests the *prior* weight, under the same binary-readout proxy flagged in A.5 — the caveat there applies here without repeating it. The framework's signature claim rests on both, and the contingency-learning literature reports the cue-density effect as the smaller and less reliably obtained of the two density effects, so the cue weight is the one whose detectability should not be assumed. **Verified numerically, for the same proxy readout** (identical design, estimator and interval construction; only the swept parameter differs, so the two tables are directly comparable; $\rho_{pr}$ held at the planning truth 0.70):

| True $\rho_{ev}$ | N = 15 | N = 30 | N = 60 | N = 120 |
|---|---|---|---|---|
| 0.20 | **1.00** | 1.00 | 1.00 | 1.00 |
| 0.40 | 1.00 | 1.00 | 1.00 | 1.00 |
| 0.60 | 0.89 | 0.99 | 1.00 | 1.00 |
| 0.80 | 0.34 | 0.60 | 0.90 | 0.99 |
| 1.00 (calibrated) | 0.04 | 0.02 | 0.03 | 0.03 |

Three things follow under the same proxy. **At equal distance from calibration the two weights cost about the same** — $\rho_{ev}$ = 0.80 gives 0.60 at thirty subjects against 0.65 for $\rho_{pr}$ = 0.80 — so there is no hidden penalty on the cue axis. **But the planned truths are not equidistant**: the planning calculation assumes $\rho_{ev} \approx 0.20$ and $\rho_{pr} \approx 0.70$, a design assumption rather than an estimate, and at $\rho_{ev} = 0.20$ the design is already at ceiling with *fifteen* subjects. Under that assumption the cue weight is the easier of the two to establish, not the harder, and the prior weight sets the sample size. **The assumption sits off this paper's own data.** At zero contingency the outcome slope over the cue slope is $\rho_{pr}/(1 - \rho_{ev})$ (§6). The assumed pair puts that ratio at 0.875, below 1, while §6's fit to Shankar et al. puts it at about 2.2 to 2.5, with even the worst case above 1. Holding $\rho_{pr}$ at 0.70, the fit places $\rho_{ev}$ near 0.7, between the 0.60 and 0.80 rows above, where power at thirty subjects falls from 0.99 to 0.60. Shankar et al. asked about a treatment's effectiveness and the planned study asks for a probability, and §2 notes the cue weight can differ between the two questions; even so, the planning calculation should be redone at the fitted ratio before a study is run. **The caveat is conditional**: this is power given the effect is as large as the framework expects. If the literature's "less robust" cue-density effect means $\rho_{ev}$ in fact sits nearer 0.80, the design needs about sixty subjects for 90% power — twice the planned thirty — and the false-exclusion rate (2–4% against a nominal 2.5%) is consistent with the nominal rate, subject to the same 200-replicate Monte Carlo uncertainty flagged in A.5.

### A.7 An exact curvature identity for the bounded readout, not an approximation

**Proved.** §6's curvature account of the Blanco, Matute & Vadillo (2013) super-additivity result rests on a qualitative claim: a bounded readout curves, so additive pushes read as super-additive below the scale's midpoint and sub-additive above it. That claim has an exact form. Let the observed mean be $\mu = h(\eta)$, for a twice-differentiable link h and an additive latent predictor $\eta$. Let a cue-side push add u to $\eta$ and an outcome-side push add v. The observed interaction — the excess of the both-pushes cell over the sum of the two one-push cells and the baseline — is

$$I(\eta;\ u,\ v) = h(\eta+u+v) - h(\eta+u) - h(\eta+v) + h(\eta).$$

Two applications of the fundamental theorem of calculus give this exactly, not as a Taylor approximation:

$$I(\eta;\ u,\ v) = \int_0^u \int_0^v h''(\eta+r+t)\, dt\, dr.$$

The 0–100 rating reads that latent predictor through a logistic curve stretched to the scale's length, $h(\eta) = L\cdot\sigma(\eta)$, with $\sigma$ the standard logistic and L = 100 rating points — the form Figure 12 already draws. Differentiating twice:

$$h''(\eta) = L\cdot\sigma(\eta)[1 - \sigma(\eta)][1 - 2\sigma(\eta)] = h(\eta)[1 - h(\eta)/L][1 - 2h(\eta)/L].$$

In words: the curvature is positive below the scale's midpoint — a mean rating under 50, which is $\eta < 0$ — and negative above it, and the inflection sits at half the scale whatever the scale's length. Carrying L rather than fixing it at 1 is what keeps the identity in the units the experiment reports: the familiar $h(1 - h)(1 - 2h)$ is the $L = 1$ case, and a number computed there is a hundred times too small to be read as a rating. Curvature alone does not fix the sign of I, because the double integral is *signed*: reversing one of the two pushes reverses I. The condition that closes the gap is that u and v share a

sign — **sufficient**, and not claimed necessary — and it is the condition §6 works under, since at null contingency both density manipulations move η the same way whenever $\rho_{ev} < 1$ and $\rho_{pr} > 0$. Under it: if the whole rectangle spanned by η, η+u, η+v and η+u+v sits below the midpoint, I is positive (super-additive); if the whole rectangle sits above it, I is negative (sub-additive); if the rectangle straddles the midpoint, curvature alone does not determine the sign and it has to be checked cell by cell. With u and v of *opposite* signs the rule inverts rather than lapsing: η = −2, u = +0.5, v = −0.5 spans a rectangle lying entirely below the midpoint and returns I = −1.99 rating points — a sub-additive interaction of about two points on the 0–100 scale, which is −0.0199 before the scale stretch. This is the identity behind the qualitative rule in §6, and it is what lets that section say precisely that Blanco et al.'s cells straddle the midpoint or sit above it rather than sit below it — a computable claim, not a visual one. *(Verified numerically: the identity to $3\times10^{-16}$, and the same-sign restriction by exhaustive sampling — of rectangles lying wholly below the midpoint, every same-signed pair gives I > 0 and every opposite-signed pair gives I < 0.)*

### A.8 The design geometry, and what the best design would look like

A.3 builds one design that works and A.4 says when a design works at all. Neither says which design is *best*, or what the cheapest working one gives up. This section answers both: the three standard optimality criteria do not compete here, and the cheapest identifying design is four cells — but four cells cost the paper the one thing §6 says it wants.

**The feasible designs form a wedge, and A.4's two structural failures are its two faces.** Work in logarithms, writing u = log s, v = log q and w = log P(C). The experimenter sets the hit rate, the false-alarm rate and the outcome's frequency; the three regressors follow. That map inverts in closed form:

```
s′ = ( e^w − e^(u+v) ) / ( 1 − e^v ).
```

In words: the false-alarm rate is whatever is left of the cue's overall frequency once the outcome-present trials have been accounted for. Substituting it back into P(C) = s·q + s′(1 − q) returns w exactly. The requirement that the cue be real and predictive — 0 < s′ < s < 1 and 0 < q < 1 — then becomes three linear inequalities in the transformed coordinates, **necessary and sufficient**:

```
u < 0,   v < 0,   u + v < w < u.
```

The cue cannot be rarer than the co-occurrence it contains, and cannot be commoner than its own hit rate. Both faces of that wedge are already in A.4's failure table. The lower face w = u + v *is* the perfectly specific cue, s′ → 0; the upper face w = u *is* null contingency, s′ = s. The two faces bound the feasible region, which is why a design pushed toward maximum spread lands on one of them — and why A.3's contour, which holds P(C) fixed, is a slice through this wedge rather than a separate construction.

**Identification is a condition on a covariance matrix, and it is computable from the planned cells.** Project out the free intercept and write C for the covariance of (u, v, w) across the design's cells, standardized so each regressor has unit variance. A Schur complement on the design's moment matrix gives rank(M) = 1 + rank(C), so A.4's rank-4 requirement says exactly this: **the design must have positive variance in every direction of the transformed regressors once the intercept is removed**, together with $\gamma \neq 0$. A.4 states the same condition; this is the form a designer can evaluate before running anyone.

**The best possible design is the orthogonal one, and it is best on all three standard criteria at once.** A designer choosing among candidate cell sets has three standard criteria, and on most problems they trade against each other: E-optimality maximizes the smallest eigenvalue of C, the worst-determined direction; D-optimality maximizes its determinant, the volume of information; A-optimality minimizes the trace of $C^{-1}$, the average coefficient variance. With unit-variance columns, $\lambda_{min}(C) \leq 1$, $\det C \leq 1$ and $\mathrm{tr}(C^{-1}) \geq 3$ — from the trace, Hadamard and harmonic-mean inequalities in turn — and **all three bounds are attained together, and only, at C = I**. There is nothing to trade. An orthogonal design in the transformed regressors is simultaneously E-, D- and A-optimal, and no numerics are needed to see it.

**Such a design fits inside the wedge, and four cells identify.** Take two levels of each transformed regressor, at a common ratio r about a centre $(\bar{s}, \bar{q}, \bar{P})$ — a multiplicative cube in the original quantities. It lies wholly inside the wedge when both faces clear **and the cube's own top hit rate is still a probability**, which takes three conditions and not one: $r^3 < \bar{P}/(\bar{s}\cdot\bar{q})$ from the lower face, $r^2 < \bar{s}/\bar{P}$ from the upper, and $\bar{s}\cdot r < 1$ so that the high level of the hit rate does not run past 1. Any of the three can bind first, so all three must be checked; the matching condition on the outcome frequency, $\bar{q}\cdot r < 1$, needs no separate check, because multiplying the two face conditions gives $r^5\cdot\bar{q} < 1$ and hence $\bar{q}\cdot r < 1$ already. Such a cube gives C = I exactly, and so does its regular half-fraction — the four cells whose three coded levels multiply to +1. Four is **support-minimal**: the model has four parameters (the intercept, the gain γ, and the two weights), and four parameters need four independent support points.

**The price of four cells is falsifiability, and that is the part worth stating rather than the saving.** In that half-fraction the two-factor interaction of the outcome-frequency and cue-marginal factors is *exactly aliased* with the hit-rate main effect. An interaction the additive model omits therefore lands entirely in a main effect, invisible rather than merely inflating a standard error: with a true main effect of 1.0 and an omitted interaction of 0.7, the four-cell design returns 1.7 and the full eight-cell cube returns 1.0. **The full eight cells buy falsifiability, not identifiability.** That is the same distinction §6 draws when it makes weight-stability across the strength cells a stated adequacy criterion rather than a nuisance — and it means identification, estimation and falsification do not share a single optimal design, so a report that gives one efficiency number has not said which of the three it means.

**A number for the identifiability cliff.** Define the margin $m = \lambda_{min}(C)$. By Weyl's inequality, every perturbation of the design's covariance smaller in norm than m leaves the design identified, and the variance in the worst-determined direction grows like 1/m. A.4's warning that a highly specific cue "sits near an identifiability cliff" is therefore computable in advance rather than qualitative: the cliff is $m \to 0$, and any planned design's distance from it is a number a referee can check.

**Prefer the smallest eigenvalue to the determinant as the first diagnostic, because losing identification *is* $\lambda_{min} \to 0$.** Drop one corner of the eight and keep equal allocation: the determinant criterion still reports 0.948 of the optimum while the smallest-eigenvalue criterion falls to 0.653. A design can look 95% efficient by volume having lost a third of its weakest identifying direction. Those two figures come from the eigenvalues 32/49, 8/7, 8/7 of the intercept-projected covariance **with the columns left unrescaled**; rescale them to unit variance and the same design reads 0.667 and 0.968 instead. Each figure has to name its matrix, for the same reason as in A.4.

**The three hypotheses are linear contrasts, so the primary test needs no ratio inference.** Write the cell mean without imposing the model's structure, $\mu = \alpha + \beta_s \log s + \beta_q \log q + \beta_C \log P(C)$. Matching this to A.4's form gives $\beta_s = \gamma$, $\beta_q = \gamma\cdot\rho_{pr}$ and $\beta_C = -\gamma\cdot\rho_{ev}$, and then, for $\gamma \neq 0$:

$$\rho_{pr} = \rho_{ev} \iff \beta_q + \beta_C = 0;\ \rho_{pr} = 1 \iff \beta_q - \beta_s = 0;\ \rho_{ev} = 1 \iff \beta_C + \beta_s = 0.$$

In words: the paper's three inferential questions — do the two corrections need weights of different sizes, is the prior correction applied in full, is the cue correction applied in full — are each a single linear restriction on coefficients the design estimates directly. The condition $\gamma \neq 0$ is **necessary**: at $\gamma = 0$ the first contrast vanishes even when the two weights differ outright. For the one-sided readings — a positive contrast meaning under-correction rather than over-correction — the sign convention $\gamma > 0$ of §2 is needed as well, and is **sufficient**.

The linear-contrast form narrows A.5's caution rather than contradicting it. The common-weight test of §6 does not have to be run on the ratio $\rho_{pr}/\rho_{ev}$ at all: the variance of a linear contrast is the ordinary $\sigma^2\cdot a^T(X^TX)^{-1}a$, with a picking out the two coefficients being added, so the skew that makes a ratio interval misbehave never enters. A Fieller-type interval is needed where a *weight* is reported as a ratio, not for the primary hypothesis.

**A limit, not a caveat: what this does not establish.** Everything above is the homoskedastic linear model with fixed column norms. §6's readout is a bounded 0–100 rating with a nonlinear link, so the apparatus-specific optimum should maximize Fisher information under that link rather than the unweighted design matrix, and that calculation has not been done here. Two consequences follow and neither is cosmetic. The thirteen-cell design of §6 is **a realistic apparatus-specific identifying design, not a proved optimum** — A.4's conditioning figures say it is well conditioned, which is a different claim. And equal allocation across its thirteen cells is a model-checking allocation rather than an estimation-optimal one, so the recommendation this section reaches is to keep all thirteen cells and stop spreading trials evenly over them, never to use fewer. The criteria named above are the standard ones of optimal experimental design; this paper does not restate their foundational literature, and the reader should not read the E, D and A labels as this paper's own coinage.

*Verified numerically:* the inverse map by back-substitution (residual $2 \times 10^{-16}$); both wedge faces against the two degenerate rates; $C = I$ to $2 \times 10^{-16}$ for the eight-cell cube and for the four-cell fraction, with all three optimality bounds attained; the aliasing by recovering 1.7 against 1.0 under an omitted interaction; and the three feasibility conditions by exhaustive scan over centres and ratios, which returns no design that satisfies all three and is nevertheless infeasible, and which exhibits centres at which each of the three binds first. Controls: duplicating one cell of the cube destroys C = I; correlating two columns moves all three criteria off their bounds together; a cube satisfying only the first feasibility condition is infeasible at the centre where the second binds; and dropping the third condition admits cubes whose high hit-rate level exceeds 1 while both face conditions pass.

## Declarations

- **Competing interests.** The author declares no competing interests.
- **Funding.** This research received no specific grant from any funding agency, commercial or not-for-profit.
- **Author contributions (Contributor Roles Taxonomy, CRediT).** A. Y. Shavit: conceptualization, methodology, formal analysis, writing — original draft, and writing — review & editing.

- **ORCID.** A. Y. Shavit — https://orcid.org/0009-0008-1235-0995

- **Use of AI and software tools.** I conceived and directed this work and am its author. Between July 2026 and October 2026, ChatGPT (OpenAI) and Claude (Anthropic) helped me search the literature, test the argument and mathematics, and write and run code, and suggested draft wording. Perplexity helped check citations, and Wolfram Mathematica helped check derivations. I did not inspect every line of generated code; deposited code, re-derivations and controls back the results. I checked every suggestion I adopted. No AI tool is an author; I take full responsibility for the content.

## Data and code availability

The model derivations and the verification code for this paper are archived at the Open Science Framework (OSF): https://osf.io/9qvbr/. That deposit contains the verification scripts that reproduce every formal claim reported here, together with their captured output. It contains no figures and no figure-drawing code, because the figures are in this paper. It contains no data: this paper collects none, and the scripts that check the formal claims generate their own designs. It also contains the two-weight fit to the published density conditions of Shankar et al. (2025) reported in §6, with its robustness and replication checks and their captured output; those scripts carry the four published condition means they use inline and redistribute no dataset and no participant-level data. It also contains the companion paper's check behind the three numbers Figure 10 prints, the early and late cross-condition contrasts and the training time at which the contrast changes sign; that script carries the colour–flavour study's per-condition trial counts inline and no participant responses. The colour–flavour identification dataset reanalysed in the companion paper (Shavit, in preparation) will be deposited with that paper.

## Preregistration

The crossed rating experiment specified in §6 will be preregistered — design, sampling plan, and the two-coefficient-separation analysis (H0: $\rho_{ev} = \rho_{pr}$) — on OSF before data collection; the power analysis behind the design is in Appendix A.5–A.6 here, and the estimator is developed in the companion paper (Shavit, in preparation).

---

---

> *We watched the cue we trained to fear,*
> *blind to the outcome's shadow trail;*
> *now name the second, weigh them both,*
> *and taste how much the sweet sign's worth.*